\documentclass[10pt]{amsart}
\usepackage[textsize=tiny]{todonotes}

\usepackage{amsmath, amsthm, amssymb}
\usepackage{graphicx}
\usepackage{varioref}

\usepackage{color}
\usepackage[dvips,left=1in, right=1in, top=1.2in, bottom=1in]{geometry}
\newcounter{notenum}

\newcommand{\argmax}{\operatorname{argmax}}

\providecommand{\R}{} \renewcommand{\R}{{\mathbb R}}

\newcommand{\N}{{\mathbb N}}
\newcommand{\PP}{{\mathbb P}}

\newcommand{\EE}{{\mathbb E}}
\newcommand{\FF}{{\mathcal F}}

\newcommand{\BB}{{\mathcal B}}

\newcommand{\UU}{\mathbb{U}}

\newcommand{\KK}{{\mathbf K}}

\newcommand{\ZZ}{{\mathcal Z}}
\newcommand{\CC}{{\mathbf C}}
\newcommand{\EN}{{\mathcal E}}

\renewcommand{\aa}{{\mathbf a}}
\newcommand{\AAA}{{\mathbb A}}
\newcommand{\A}{{\mathcal A}}

\newcommand{\EEE}{\mathbf{E}}

\newcommand{\el}{{\mathbb L}} 
\newcommand{\lzer}{\el^0}

\newcommand{\ltwo}{\el^2}

\newcommand{\ini}{\mathbf c}

\newcommand{\nC}{\tilde{C}}
\newcommand{\nG}{\tilde{G}}

\newcommand{\npi}{\tilde{\pi}}

\newcommand{\npp}{\tilde{\pp}}

\newcommand{\naa}{\tilde{\aa}}

\newcommand{\Br}{B^{\mathsf{r}}}
\newcommand{\br}{b^{\mathsf{r}}}
\newcommand{\Zr}{Z^{\mathsf{r}}}
\newcommand{\aar}{\aa^{\mathsf{r}}}
\newcommand{\Cr}{C^{\mathsf{r}}}
\newcommand{\ppr}{\pp^{\mathsf{r}}}
\newcommand{\pir}{\pi^{\mathsf{r}}}

\newcommand{\Bn}{B^{\diamond}}
\newcommand{\BBn}{\BB^{\diamond}}
\newcommand{\Zn}{Z^{\diamond}}

\newcommand{\Cn}{C^{\diamond}}
\newcommand{\ppn}{\pp^{\diamond}}
\newcommand{\pn}{p^{\diamond}}
\newcommand{\pin}{\pi^{\diamond}}
\newcommand{\bn}{b^{\diamond}}

\newcommand{\nBr}{\tilde{B}^{\mathsf{r}}}
\newcommand{\nZr}{\tilde{Z}^{\mathsf{r}}}

\newcommand{\nCr}{\tilde{C}^{\mathsf{r}}}
\newcommand{\nppr}{\tilde{\pp}^{\mathsf{r}}}

\newcommand{\nBn}{\tilde{B}^{\diamond}}
\newcommand{\nBBn}{\tilde{\BB}^{\diamond}}
\newcommand{\nZn}{\tilde{Z}^{\diamond}}

\newcommand{\nCn}{\tilde{C}^{\diamond}}
\newcommand{\nppn}{\tilde{\pp}^{\diamond}}

\newcommand{\npin}{\tilde{\pi}^{\diamond}}

\newcommand{\dpp}{\hat{\pp}^{*}}
\newcommand{\daa}{\hat{\aa}^{*}}

\newcommand{\dZr}{\hat{Z}^{\mathsf{r}}}

\newcommand{\dppr}{\hat{\pp}^{\mathsf{r}}}

\newcommand{\inir}{\ini^{\mathsf{r}}}

\newcommand{\dZn}{\hat{Z}^{\diamond}}

\newcommand{\dppn}{\hat{\pp}^{\diamond}}

\newcommand{\inin}{\ini^{\diamond}}

\newcommand{\gi}{\gamma_i}
\newcommand{\Var}{\operatorname{\mathbb{V}ar}}
\newcommand{\spa}{\operatorname{span}}
\newcommand{\Cov}{\operatorname{\mathbb{C}ov}}
\newcommand{\pp}{\mathbf{p}}
\newcommand{\ppi}{\mathbf{\pi}}

\newtheorem{theorem}{Theorem}[section]
\theoremstyle{plain}

\newtheorem{standing assumption}[theorem]{Standing Assumption}

\newtheorem{corollary}[theorem]{Corollary}
\newtheorem{definition}[theorem]{Definition}

\newtheorem{proposition}[theorem]{Proposition}
\newtheorem{remark}[theorem]{Remark}

\numberwithin{equation}{section}
\begin{document}

\title{\textbf{The Effect of Market Power on Risk-Sharing}}

\maketitle

\begin{center}
\begin{minipage}{0.8\textwidth}
\begin{center}
{\bf\large Michail Anthropelos$^{*,\dag}$}\\
{\tt anthropel@unipi.gr}\\
\end{center}
\end{minipage}
\end{center}
\ \\[0.2ex]

\begin{quote}
\noindent{\bf Abstract.} The paper studies an oligopolistic
equilibrium model of financial agents who aim to share their
random endowments. The risk-sharing securities
and their prices are endogenously determined as the outcome of a
strategic game played among all the participating agents. In the
complete-market setting, each agent's set of strategic choices
consists of the security payoffs and the pricing kernel that are consistent with the
optimal-sharing rules; while in the incomplete setting, agents
respond via demand functions on a vector of given tradeable
securities. It is shown that at the (Nash) risk-sharing
equilibrium, the sharing securities are suboptimal, since agents submit for sharing different risk exposures than their true endowments. On the other hand, the Nash equilibrium prices stay unaffected by the game only in the special case of agents with the
same risk aversion. In addition, agents with sufficiently lower risk aversion act as predatory traders, since they absorb utility surplus from the high risk averse agents and reduce
the efficiency of sharing. The main results of the paper also hold under the generalized models that allow the presence of noise traders and heterogeneity in agents' beliefs. \footnote{$^*$I would like to thank Peter
Bank, Jak\v{s}a Cvitani\'{c}, Paolo Guasoni, Constantinos
Kardaras, Rohit Rahi, George Skiadopoulos, Herakles Polemarchakis,
Dimitri Vayanos, Dimitris Voliotis, Gordan \v{Z}itkovi\'{c} and
the seminar and conference participants in Technische
Universit\"{a}t in Berlin, London School of Economics, Columbia
University, University of Texas at Austin, Leibniz Universit\"{a}t
in Hannover and University of
Piraeus. I also thank the anonymous referees for their valuable comments and suggestions that have improved this paper.}\footnote{$^{\dag}$Department of Banking and Financial
Management, University of Piraeus, Karaoli and Dimitriou Str 80,
Piraeus, 18534, Greece; Tel.~+30-2104142551; Fax.~+30-2104142331.}
\end{quote}
\

\noindent{\bf Keywords:} Optimal risk-sharing, Nash equilibrium in
risk-sharing, security designing, predatory trading, thin markets.\

\noindent{\bf JEL Classification:} D53, D43, C72.\\

\section{Introduction}\label{sec:intro}

The concept of risk sharing is central to a large variety of
financial fields ranging from investment management and structured
finance to insurance and derivative markets. Its importance stems
from the fact that investors, financial institutions and insurers
find that the sharing of investment payoffs, defaultable
incomes and insurance liabilities is often mutually beneficial in
terms of reduction of risk exposures. The search for the best way
to share risk is connected to financial innovation, in the sense
that such sharing is fulfilled by designing and trading new
financial securities.

The large majority of the growing body of research in risk-sharing
assumes that agents act as \textit{perfect competitors} and the
induced allocation of risk is Pareto optimal. However, when financial agents negotiate the sharing of their
(otherwise unhedgeable) risk exposures by designing new
or by trading given securities, participation is normally
limited (at least at the primary market level) to some
institutions and/or some of their clients.$^1$\footnote{$^1$Although the large participation in the majority of financial markets does not leave room to large agents for market
manipulation, several empirical studies have argued that in
financial transactions among institutional investors some form of
market power is used (see among others Chan and Lakonishok \cite{ChaLak95}, Keim and Madhavan \cite{KeiMad95,
KeiMad97} and  Kraus and Stoll \cite{KraSto72}). Also, other empirical evidence indicates that in some cases even
market makers had acted strategically in placing their
orders (see e.g., Christie et al.~\cite{ChrHarSch94}, Christie and Schultz \cite{ChrSch94}, Gibson et al.~\cite{GibSinYer03} and the 
the related discussion by Liu and Wang in \cite{LiWan16}). These studies have shown that the market
power did exist, was exploited and did cause inefficient
allocations and prices.} This limited participation implies that at least some of the agents possess the
power to affect the equilibrium allocations and prices. As discussed for instance in Rostek and Weretka \cite{RosWer15}, even in the contemporaneous financial markets, large institutional investors often have the power to dominate the trading and drive the prices according to their own benefit. Moreover, when the market is over-the-counter (OTC), the assumption
of perfectly competitive structure is even further away from being
realistic. An indicative example is the reinsurance market, where few insurance companies 
share their portfolios through the trading of new reinsurance contracts. It is apparent that such market can not be considered competitive, since the strategic behavior of each individual company could heavily influence the equilibrium transaction.

Market models that do not take into
account agents' market power lead to equilibria that \textit{overestimate}
the market efficiency (and, as stated by Hellwig in \cite{Hel80} and
Kyle in \cite{Kyle89}, they require agents to behave ``schizophrenically''
by ignoring their ability to affect the market). If we assume that
participating agents do exploit their impact on the way the risk
is shared, the equilibrium is different and the efficiency is most
likely reduced.

Following the standard setting in the risk-sharing literature, we
assume that agents are endowed with some risky portfolios, which are called \textit{random endowments}. The mutually
beneficial sharing of these endowments can be achieved in a complete
or in a incomplete-market setting. In the former, agents design and
price new securities that Pareto optimally share their risk exposures.$^2$\footnote{$^2$The securities that share risk in a
Pareto optimal way were introduced by the seminal works of Borch 
\cite{Bor62, Bor68} and Wilson \cite{Wil68} (see also Demange and Laroque \cite{DemLar95} and
the surveys of Allen and Gale in \cite{AllGal94} and Duffie and Rahi in \cite{DuffRah95}).} On the other
hand, if the securitization of the agents' endowments is not
possible (due to exogenous constrains, such as transaction costs
and strict regulation), an \textit{incomplete} risk-sharing can be attained through the trading of
 a \textit{given} vector of securities. Although
there is an extended literature dealing with both settings, the effect of agents' market power on equilibria has not been sufficiently addressed. The main
objective of this paper is to fill this gap by (i) modelling the
strategic behavior of agents with heterogenous risk aversions when
they share risk, (ii) establishing and analyzing the Nash
equilibria that occur as the outcome of these strategies and (iii)
investigating which are the agents (if any) that benefit from the
oligopolistic structure of a risk-sharing transaction.

\subsection{Model description and main findings}

We consider a \textit{one-shot}, risk-sharing transaction among
risk averse agents. If none of them acts strategically, their endowments are shared through the optimal sharing rules; the risk-sharing securities and the pricing kernel are given by specific functions of agents' endowments, in a way that the aggregate utility after the transaction is maximized. 

In our model, we suppose that agents agree that whatever vector of endowments is submitted for sharing, is going to be shared according to the optimal sharing rules. We then take the position of an individual agent and ask whether she has motive to submit risk exposure different
than her true one. Since the risk-sharing securities and the pricing kernel are functions
of the submitted endowments, the endowment that she chooses to
submit directly affects not only the structure of the designed securities, but also their prices.
According to the proposed model, she should respond as her
to-be-shared endowment the random quantity (called \textit{best-endowment response}) that maximizes her own utility, when the
agreed sharing rules are applied.$^{3,4}$\footnote{$^{3}$In
contrast to the relevant literature on thin markets with symmetric information (see
e.g., Weretka \cite{Wer11} and the references therein), here the strategic
set of choices is of infinite dimension; namely it is the set of
all random variables that can be considered as agent's random
endowment. This set of choices is in
fact equivalent to the set of securities and pricing kernel that are
consistent with the optimal sharing rules.}\footnote{$^{4}$The
generality of the agent's random endowment in our model allows
its application to a number of financial
markets, such as the trading of innovated derivative products,
the use of reinsurance contracts or the transactions among
inter-dealers (see also the related discussion in
Vayanos \cite{Vay99}).} Under mean-variance preferences (hereafter M-V preferences),
it is shown that it is \textit{never optimal} for an agent to
submit for sharing her true risk exposure. Instead, each agent has motive to declare only a fraction of her true endowment and also to report exposure to the aggregate endowment submitted by the other agents. 

A similar best-response strategy is employed when agents negotiate
the trading of a given bundle of tradeable securities
(incomplete-market setting). Therein, we consider an individual
agent who knows the aggregate demand schedule of the rest of the
agents$^{5}$\footnote{$^{5}$Note that under M-V preferences, the demand schedules are linear and hence a demand function can be inferred by knowing only two of its values (orders). This implies that for the implementation of the proposed strategy (and the induced Nash-game), agents do not need to submit their entire demand functions, but rather only two of its values. This makes the model more practical and applicable.} and ask:
\textit{Which is her best equilibrium price, given the submitted demand from the rest of the agents?} Naturally,
we define as her ``best equilibrium price'' the one that clears out the market and at the same time maximizes her own utility. She can then drive the market to
this price by submitting a corresponding demand function (called \textit{best-demand response}). As explained in the sequel, this best-response strategy is appropriate when the transaction involves risk-sharing, since it allows us to identify the hedging needs that are revealed by the submitted demand functions. It is shown that
under M-V preferences, the best-demand response is
different than the true one in any non-trivial case
(i.e., in any case where some risk is to be shared). In fact, the
best-demand response coincides with the demand that would be asked by an agent endowed with
the best-endowment response.

The aforementioned strategic behavior, when applied by all agents,
forms a negotiation scheme on the risk-sharing transactions. The risk-sharing
equilibrium is the outcome of a Nash-type game played among the
participating agents, where their sets of strategic choices are the
risks they choose to submit (or equivalently the securities they
are willing to get). In Section \ref{sec:nash equilibria}, it is
proved that under M-V preferences the complete-market setting
admits a pure-strategy Nash equilibrium. Furthermore, although it
is rare to have uniqueness of Nash equilibrium in models with
uncertainty (see among others, Grossman \cite{Gro81} and Klemperer and Meyer \cite{KleMey89}),
the equilibria are indeed \textit{unique} and fully characterized.
In the incomplete market, the game is played on
the agents' reported demand schedules or equivalently on the
prices and the allocations of the tradeable assets. The Nash equilibrium is again unique and given in explicit form.

In the Nash risk-sharing, the equilibrium is inefficient in all
non-trivial cases, meaning that the aggregate utility is reduced when compared to the optimal one. Interestingly enough though, for agents with sufficiently lower risk
aversion the Nash equilibrium yields higher utility. The main
message of this theoretical result is that \textit{in thin risk-sharing markets, not only there is a loss in the aggregate utility, but also agents with relatively lower
risk aversion absorb utility surpluses from their counterparties}.$^{6}$\footnote{$^{6}$The comparison of utility surpluses in competitive and in Nash equilibria is based on the simplified (but indicative) market of two agents with different risk aversions.} Another way to interpret this result is through the notion of \textit{predatory trading} (introduced by Brunnermeier and Pedersen in \cite{BruPed05}). More precisely, the market's thinness gives an opportunity for the relatively lower risk averse agents to act as predatory traders and exploit the intense hedging needs of their counterparties. The strategic behavior is more effective for the more risk tolerant agents, who in a sense ``dominate'' the transaction. In contrast to the other predatory trading models, this is an endogenous result and holds even if predatory traders are still risk averse.$^7$\footnote{$^7$The fact that the lower risk averse agents tend to get more utility in Nash equilibrium is heavily based on the generality of the strategic set and in particular on the agents' ability to declare as their risk exposure any random variable on the probability space. One may ask whether similar result holds when the set of strategic choices is exogenously restricted. In Appendix \ref{sec:best response with percentage}, we consider an example of such restriction where agents choose only the size of their true random endowments that are willing to share. Through this indicative (counter)example, it is shown that the answer is negative, meaning that in complete-market setting the effectiveness of the predatory trading strongly depends on the imposed set of strategic choices.} Furthermore, the securities
that are designed to share risk are the optimal ones only under
the special case of agents with the same risk aversion. Even in
this case however, the equilibrium volume and the
efficiency are reduced.

In the incomplete-market setting, the induced Nash equilibrium can be
considered as an \textit{oligopoly variation of the CAPM}. In
Section \ref{sec:nash equilibria}, it is shown that \textit{the
Nash equilibrium prices are equal to the perfect-competition ones
if and only if agents are homogeneous with respect to their risk
preferences}. Even in this case however, the allocation in Nash
equilibrium is not efficient and the volume is lower (in
particular, the volume percentage reduction is $1/n$, where $n$ is
the number of participating agents). In the more general
situations of heterogeneous agents, we establish an exact measure
of the \textit{price impact} that is caused by the agents' game. For example, it is shown that an intense upward price pressure occurs not only
when the agents with intense hedging demand are also high risk averse, but even when agents with low risk aversion (acting as predatory traders) participate in the trading. As in the complete-market setting, sufficiently lower
risk averse agents profit from the market's inefficiency, for any vector of tradeable securities.

For the definition of market inefficiency, we follow the related literature (see among others Acharya and Bisin \cite{AchaBin05}) and define the 
\textit{risk-sharing inefficiency} as the difference
between the aggregate utility surplus at the optimal
risk-sharing and the aggregate utility surplus at the Nash
equilibrium. As stated above, inefficiency is positive, whereas its
size is mainly determined by the number of participating agents and the market's completion. It is
reasonable to expect that the efficiency of the risk-sharing
transaction increases as the number of the participating
agents increases. Indeed, both in complete and incomplete-market
settings, the differences between the Nash and competitive
equilibria vanish and the market becomes efficient as the number
of agents increases to infinity. In addition, a market that is closer to
completion leans to more efficient sharing. It is shown that even when the market is
thin and gives rise to market power, each agent gets more utility when the
market's setting is complete. This implies that even in
markets with limited participation, financial innovation is mutually beneficial (at least when there are no further transaction costs).

For the aforementioned results, it has been assumed that the net supply of risk is zero. This is basically because there is no noise traders in the risk-sharing scheme who could submit specific endowment or demand orders. In Section \ref{sec: noise trader}, we generalize our settings and examine whether and how noise traders change the equilibria and the consequent results. For this, we introduce the noise traders' aggregate orders following the lines of the classic model of Kyle  \cite{Kyle89} (see also Rostek and Weretka \cite{RosWer15}). In particular, noise traders submit an aggregate endowment that they want sell, say $\EN_N$, without actually get any long position on agents' endowments; and the price they get from selling $\EN_N$ is determined as part of the equilibrium.  Similarly, in the incomplete-market setting, noise traders submit a given and fixed order on the tradeable securities that is to be satisfied at any price level. It turns out that the main messages of the paper stay unaffected by the presence of noise traders. For the best-response strategies, each agent faces the other agents' aggregate risk or demand plus the noise traders' fixed order. This implies that she has motive to submit exposure the noise traders' risk too, which hereto changes its allocation in the equilibrium. In particular, low risk averse agents buy less of endowment $\EN_N$ at the Nash equilibrium, which eventually yields again higher utility than in the optimal sharing. One noticeable observation is that the predatory trading of the low risk averse agents does not necessarily become more profitable by the presence of noise traders. We show, for example, that when $\EN_N$ is positively correlated with a predatory trader's endowment, the noise traders' order reduces her utility, both in competitive and Nash equilibrium. However, for sufficiently lower risk aversion, the surplus at Nash equilibrium is still higher, in any non-trivial case. 

Another possible generalization of the model is regarding agents' beliefs. As in the majority of the relevant literature, we have assumed that agents' probability assessments for payoffs of tradeable securities and endowments are common. In Section \ref{sec:difference}, we withdraw this assumption in the incomplete-market setting and study how this affects the corresponding equilibria. We introduce this generalization following the recently developed models of Kyle et al.~\cite{KylObiWan14} (see also Rostek and Weretka \cite{RosWer12}) and show that the way agents share the deviations of their beliefs is similar to the way they share their random endowments both in competitive and Nash equilibrium. In particular, except from revealing different hedging needs, agents have motive to declare different beliefs on the tradeable securities too. Homogeneity in agents' risk aversion keep the equilibrium prices unaffected by the game, but the corresponding volume is again reduced. More interesting is the discussion regarding the utility surpluses. At first glance, one could conjecture that the difference in beliefs results in a more profitable predatory trading, since, apart from the hedging needs, a predatory trader could exploit the deviation on probability assessments. However, this is not necessarily true. In our model, predatory traders are risk averse with specific hedging needs. This means that when the average beliefs of the other agents induce higher (resp.~lower) value for the securities a predatory trader buys (resp.~sells), her utility surpluses are afflicted. Nevertheless, even under the model that allows different beliefs, sufficiently lower risk averse agents still get more utility at Nash equilibrium than at the competitive one.


\subsection{Connection with the relevant literature}

The paper is related to two main strands of literature: the design of financial securities (related to the complete-market setting) and the trading of standardized securities in thin (OTC) markets under imperfect competition (related to the incomplete-market setting). The contributions of the paper and its connections to these strands of the literature are described below. 

The large majority of the literature on financial innovation and more particularly on security design imposes a competitive market structure and does not allow agents to implement any kind of strategic behavior (see among others Acciaio \cite{Acc07}, Barrieu and El Karoui \cite{ElKBarr05}, Dana \cite{Dan11}, and Jouini et al.~\cite{JouSchTou08} and the surveys of Allen and Gale \cite{AllGal94} and Duffie and Rahi \cite{DuffRah95}). Relatively 
little attention has been paid to the more realistic risk-sharing transactions, where agents possess and exploit some sort of market power.

In the existed models of non-competitive risk-sharing, the agents' market power is \textit{asymmetric} and stems mostly from asymmetric information or other types of imposed structural differences among players. For example, games on security design under adverse-selection problem (i.e., asymmetric information) have been recently studied (see among others Horst and Moreno-Bromberg \cite{HorMorBro08, HorMorBro11} and Page and Monteiro \cite{PagMont03, PagMont07}), where the market consists of firms (security issuers) that offer a menu of contracts to a continuum of agents. In Section \ref{sec:Agent's Strategic Behavior in Risk Sharing Market}, we also consider the case where only one agent has asymmetric information on the aggregate endowment of the rest of the agents. However, the risk-sharing transaction that we study here is not between principal(s) and agent(s), but rather the security design involves a mutually beneficial transfer of all the agents' risks (agents are not categorized with respect to their types, but with respect to their endowments and risk aversions).

More recently, other types of discriminations among the participating players have been imposed in modelling of security design games. Agents are distinguished as arbitrageurs and investors (see Rahi and Zigrand \cite{RahZig09} and Shleifer and Vishny \cite{ShlVis97}) or security issuers and investors (see Carvajal et al.~\cite{CarRosWer12} and the reference therein)$^{8}$\footnote{$^{8}$Rahi and Zigrand in \cite{RahZig09} study a security-design game played among arbitrageurs, where the induced Nash equilibrium is
optimal for the arbitrageurs regarding their profits from the
mispricings across different market segments. Although the
arbitrageurs' strategic behavior has some common features with the
model presented here (e.g., it refers to maximization of quadratic
utility functions), the game is played on a different field
(arbitrageurs imperfectly compete on the profits that come as a
result of segmented markets and they do not share any risk
exposure). In Carvajal et al.~\cite{CarRosWer12} on the other hand, only some of the participating agents, namely the security issuers (entrepreneurs) design the securities and investors competitively trade them, without securitizing their own endowments.}. As already emphasized, the results of this literature do not cover our model, since we do not impose any agents' discrimination regarding their information or their ability to influence the market. To the best of our knowledge, this is the first attempt to model a symmetric game on risk-sharing transaction where the agents' bargaining power and security payoffs are endogenously determined in the equilibrium.$^{9}$\footnote{$^{9}$Another novelty is that the set of strategic choices consists of the all possible endowments that each agent can submit for sharing. This general setting allows, among other things, to conclude that even when agents impose sharing rules that are consistent with the optimal ones, the securities that they design differ from the Pareto-optimal securities, since agents do not submit their true endowments for sharing.} In particular, it is an endogenous result that sufficiently lower risk averse agents have higher bargaining power in the game regardless of their true random endowments.  
\smallskip

The game on the incomplete market-setting is related to literature on imperfectly competitive financial markets, which is based on the seminal work of Kyle in \cite{Kyle89}. According to this theory, agents act strategically on the trading of \textit{given} securities, and their strategic set of choices consists of the demand functions that they may submit. The related literature follows this approach and studies equilibria under a variety of different sources of imperfection. Mainly (as in Back et al.~\cite{BacCaoWil00}, Kyle \cite{Kyle89}, Kyle et al.~\cite{KylObiWan14} and Vayanos \cite{Vay99}), the source of market's imperfection is the asymmetric information, where agents are categorized to informed, uniformed and noise ones. Our model departs from this literature, since we assume an oligopolistic market, where \textit{each} agent is rational and strategically uses her market power even without the presence of noise traders$^{10}$\footnote{$^{10}$As already mentioned, the presence of noise traders and its impact on the equilibrium are addressed in Section \ref{sec: noise trader} as an extension of our basic model.}. In other words, as in the complete-market setting, it is the thinness of the market that gives rise to the market power, rather than any type of exogenously imposed asymmetry.  

Recently, a number of equilibrium models have been established assuming that market's imperfection stems from market's thinness. More precisely, in Carvajal and Weretka \cite{CarWer12} (see also Malamud and Rostek \cite{MalRos15},  Weretka \cite{Wer11} or Carvajal \cite{Car14} for a broader discussion), a non-competitive market without asymmetric information is considered and (as in our paper) each agent submits demand functions taking into account the impact of her order in the equilibrium. The main substantial difference to our demand-function game is the imposed set of strategic choices. 
According to the demand games in Carvajal and Weretka \cite{CarWer12}, Kyle \cite{Kyle89} and Weretka \cite{Wer11} (see also Rostek and Weretka \cite{RosWer15} for a dynamic version), each agent estimates the price impact of her order, (defined as the slope of the aggregate demand of the rest of the agents) and responds accordingly. Under linear demand functions however, the slope of the demands does not depend on agents' endowments and hence this demand game is not driven by the agents' motive to reveal hedging needs different from their true ones. In technical terms, our game is played on the intercept points of the linear demand functions rather than their slopes. As discussed in details in Section \ref{sec:nash equilibria}, it is exactly the strategic behavior on the intercept points that allows us to examine the effective sharing of risks in Nash equilibrium and compare it with the corresponding equilibrium in the complete-market setting.$^{11}$\footnote{$^{11}$Note also that although in Carvajal and Weretka \cite{CarWer12} and Weretka \cite{Wer11} the agents' preferences are more general, the tractability of their models is based on the assumption of finite probability space.} In addition, it should be emphasized at this point that the Nash-game on the intercept points has an equilibrium even in the case of two agents. This comes in sharp contrast to models of Carvajal and Weretka \cite{CarWer12}, Kyle \cite{Kyle89} and Weretka \cite{Wer11}, where the two-agent game is in fact ill-posed. This should be highlighted, since many of the real-world risk-sharing transactions are between only two agents. 

In the OTC transactions of given securities, the market power possessed by large institutional investors is more intense. In contrast to our model, existed models on OTC market valuation imposes one-side market power or different levels of bargaining power. For example, Duffie et al.~\cite{DufGArPed07} (see also Duffie et al.~\cite{DufGArPed05}) models the bilateral OTC transactions among agents who have different bargaining power, and exogenously given preference shocks determine which one is the buyer and which the seller of the tradeable securities. Other strands of the literature in the modelling of OTC transactions, focus on the impact on the equilibrium prices that is caused by market makers' actions. For example, in Liu and Wang \cite{LiWan16} (see also the references therein) it is eventually assumed that the market makers possess all the bargaining power of the game and act strategically according to their own utility optimization criteria.  
In contrast to these models, we do not postulate that certain agents act as market makers or have exogenously given higher bargaining power. Rather, it is \textit{an endogenous result} that agents with sufficiently lower risk aversion effectively end up intermediating and get more gain from the market's thinness in both complete and incomplete setting (see also Remarks \ref{rem:speculator} and \ref{rem:speculator Nash}).$^{12}$\footnote{$^{12}$The so-called strategic security market models that are based on the seminal work of Shapley and Shubik in \cite{ShaShu77} are also related to our incomplete-market setting (see among others Koutsougeras \cite{Kou03}, Koutsougeras and Papadopoulos \cite{KouPap04} and Peck and Shell  \cite{PecShel96}). In this literature however, the pre-specified allocation and pricing mechanism is not consistent with our equilibrium setting (for instance, Koutsougeras and Papadopoulos in \cite{KouPap04} impose strict budget constrains, while the prices are calculated as the ratio of the aggregate amount of money put up for each security over the aggregate asked units of the security). Under such mechanism, CAPM produces non-linear pricing, and arbitrage opportunities could arise.}

Also, another branch of the literature on market-power modelling (see e.g., Brunnermeier and Pedersen \cite{BruPed05} and Vayanos \cite{Vay01}) distinguishes agents between
price-takers and those with market power (or large strategic
investors and noisy traders$^{13}$\footnote{$^{13}$ In our model, noise traders do not change the main structure of the equilibrium. As analysed in Section \ref{sec: noise trader}, their presence can be summarized through a given submitted endowment or a given order on the tradeable securities (both considered as public information). This eventually means that the net supply, in both complete and incomplete market settings, is not zero.}). In particular, in Brunnermeier and Pedersen \cite{BruPed05}, a scheme of a \textit{predatory trading} is created by the (exogenously given) need of certain investors to liquidate their positions. The predator investor exploits this need by acting strategically and at first sells the asset and subsequently buys it back. In our model, the predatory trading is given endogenously by the strategic behavior of the agents who respond to the other agents submitted hedging needs. As mentioned above, it is shown that the role of the predator is played by agents with sufficiently lower risk aversion, who exploit the intense hedging needs of the higher risk averse counterparties. Furthermore, if an agent also possesses asymmetric information on the other agents' endowments, the predatory trading is more excruciating and more profitable. 

\smallskip

Finally, we highlight the two main features of this paper, that distinguish it from the related literature. First, the agents' endowments do not necessarily belong in the span of the tradeable securities (as in Rostek and Weretka \cite{RosWer12, RosWer15} and Vayanos \cite{Vay99} for instance). Besides that this generalization is closer to the real-world situations, it allows us to examine how a pre-existing unhedgeable payoff in an agent's portfolio affects her strategic behavior. Most importantly, assuming general endowments, we are able to designate the link between the equilibria in compete market  and incomplete market setting, and in particular that Nash equilibrium demands reflect the hedging needs from the Nash equilibrium endowments.

Secondly, in contrast to the majority of literature in risk-sharing and non-competitive markets that assumes equal agents' risk aversions, our results emphasize the importance of agents' heterogeneity on their risk preferences, not only for the allocation and prices, but mainly for the utility surplus that each individual agent gets at the equilibria.

\bigskip

The paper proceeds as follows: Section \ref{sec:Market set up}
introduces the market model and sets up the optimal risk-sharing
rules in both complete and incomplete market settings.
Section \ref{sec:Agent's Strategic Behavior in Risk Sharing
Market} establishes the model for an individual agent's strategy,
regarding the risk that she chooses to share and the demand function she chooses to submit, given the other agents' submitted risks and demands. 
The Nash risk-sharing equilibria are defined and
analyzed in Section \ref{sec:nash equilibria}. The effect on the main findings caused by the presence of noise traders is 
addressed in Section \ref{sec: noise trader}; while Section \ref{sec:difference} is dedicated to the market model that allows agents' with different beliefs. In Appendix \ref{sec:best response with percentage}, an indicative example of a restricted strategic set is discussed and the paper's contributions are summarized
in Section \ref{sec:conclusion}. For the reader's convenience, the
proofs are omitted from the main body of the paper and provided in
Appendix \ref{sec:proofs}.

\bigskip

\section{Risk-Sharing Equilibria without Market Power}\label{sec:Market set up}

We consider a static market model of $n$ agents who aim to reduce
their risk exposures by trading to each other. Throughout this
manuscript, it is assumed that there exists an exogenously priced
num\'{e}raire in units of which all mentioned payoffs are denominated. Each agent is already exposed to a
random endowment that incorporates the net
discounted payoffs of all the unhedgeable financial positions that
she has taken (the risk that cannot be hedged out by trading in
any market outside this setting). These endowments are
denoted by random variables $\EN_i$, $i\in\{1,...,n\}$, which
are defined in a standard probability space $(\Omega, \FF,
\PP)^{14}$\footnote{$^{14}\PP$ is the so-called
``\textit{subjective}'' probability measure, and assumed to be
common for each agent. In Section \ref{sec:difference}, we examine how the the main results of the paper are affected we withdraw the common-beliefs assumption.}. The sum of these
random endowments is called the \textit{aggregate endowment} and is
denoted by $\EN:=\sum_{i=1}^n\EN_i$. The preferences of agent
$i$ are modelled by the M-V utility$^{15}$\footnote{$^{15}$The quadratic utility has be widely
used in the risk-sharing literature in a variety of subjects, from
adverse-selection problems (see e.g.,
Carlier et al.~\cite{CarEkTou07}, Horst and Moreno-Bromberg \cite{HorMorBro08, HorMorBro11}) to games on financial
innovation (as in Rahi and Zigrand \cite{RahZig09}). Furthermore, these preferences are consistent with the standard assumptions of exponential utility and normally distributed payoffs (see among others Kyle \cite{Kyle89}, Vayanos \cite{Vay99} and Liu and Wang \cite{LiWan16}). Note also 
that M-V preferences can generally be considered as a second order
approximation of other utility functionals (see for instance
Section 4 of Anthropelos and \v{Z}itkovi\'{c} \cite{AnthZit10a}).}
\begin{equation}\label{eq:quadratic functional}
\UU_i(X)=\EE[X]-\gamma_i\Var[X],
\end{equation}
for any random payoff $X$ that belongs in
$\ltwo(\Omega,\FF,\PP)$ (hereafter denoted simply by $\ltwo$), where 
constant $\gamma_i>0$ is agent $i$'s \textit{risk-aversion coefficient}
and $\EE[\cdot]$ and $\Var[\cdot]$ stand for the expectation
and variance maps under probability measure $\PP$. With a slight
abuse of notation, when $\Var[\cdot]$ is applied to a vector of
random variables, it refers to the associated variance-covariance
matrix.

In the sequel, we consider two risk-sharing settings: the incomplete-market one, where agents share their risks through the trading of a \textit{given} vector of securities; and the complete-market setting where risk-sharing involves the designing and trading of new financial securities.

\subsection{Risk-sharing in incomplete markets}\label{sec:constrainted}

As discussed in the introductory section, the design of new
securities that optimally share the agents' risky endowments and complete the market is rarely
possible.$^{16}$\footnote{$^{16}$Examples of the exogenous constrains
that restrict agents from trading the optimal risk-sharing
securities are: (i) The stricter regulation on OTC
transactions, that imposes restrictions on credit levels, (ii) the
indivisibility of some types of random incomes, e.g., real estate
investments, revenues of a running business, dividends from
illiquid shares etc, (iii) further liquidity concerns and
transaction costs that make the optimal risk-sharing trading
disadvantageous (an issue addressed among others by Allen and Gale in \cite{AllGal94}).} In such situations, agents can mutually reduce
their risk exposures by trading certain number of \textit{standardized}
securities, whose payoffs are possibly correlated with their endowments. These securities (hereafter called \textit{tradeable}) could be any structured financial derivatives, such as credit derivatives, asset backed securities, reinsurance contracts etc. Although, trading a given vector of securities is not a Pareto-optimal transaction, an equilibrium allocation could improve the utility of 
each individual agent.

We assume the existence of $k\in\N$ tradeable securities, with random vector $\CC=(C_1,...,C_k)\in(\ltwo)^k$ denoting their payoffs. For any price vector $\pp\in\R^k$, agent $i$'s optimization problem is 
\begin{equation}\label{eq:utility after buying a at p}
\underset{\aa\in\R^k}{\sup}\{\UU_i(\EN_i+\aa\cdot\CC-\aa\cdot\pp)\}
=\underset{\aa\in\R^k}{\sup}\{\UU_i(\EN_i+\aa\cdot\CC)-\aa\cdot\pp\}.
\end{equation}
The set of vectors that maximize \eqref{eq:utility
after buying a at p} for a given price vector $\pp$ is the
\textit{demand} of agent $i$ on $\CC$ at price $\pp$, i.e., $Z_i(\pp):=\underset{\aa\in\R^k}{\argmax}\left\{\UU_i(\EN_i+\aa\cdot\CC)-\aa\cdot\pp\right\}^{17}$\footnote{$^{17}$Note that this optimization problem imposes no short-selling constrains. One may think that the absence of such
constrains implies the possibility of unbounded supply or demand. This can be avoided, for instance, if we impose some regularity requirements on the
set of agent's positions. Unbounded supply (demand) indeed occurs when the prices are sufficiently high (or low). However, the
equilibrium arguments, developed in the sequel, will endogenously exclude such
extreme situations.}. For the set $Z_i(\pp)$ to be a singleton for any price vector $\pp\in\R^k$, it is sufficient to impose the following assumption.

\noindent \textbf{Standing Assumption.} \textit{For every
considered vector of securities $\CC$, the matrix $\Var[\CC]$ is
non-singular}. \

This condition guarantees that $Z_i(\pp)$ is indeed a function from $\R^k$ to
$\R^k$, which is called the \textit{demand function} of agent $i$ of securities $\CC$, and under M-V preferences
\begin{equation}\label{eq:quadr demand}
    Z_i(\pp)=\left(\frac{\EE[\CC]-\pp}{2\gi}-\Cov(\CC,\EN_i)\right)\cdot\Var^{-1}[\CC],
\end{equation}
where $\EE[\CC]$ stands for the vector $(\EE[C_1],...,\EE[C_k])\in\R^k$,
and for any payoff $X\in\ltwo$, $\Cov(\CC,X)$ denotes the vector
$(\Cov(C_1,X),...,\Cov(C_k,X))\in\R^k$. 
\begin{remark}\label{rem:demand decomposition}
Note that the demand has two
distinguished sources; the risk premium: $(\EE[\CC]-\pp)/2\gi\cdot\Var^{-1}[\CC]$,
and the correlation between the tradeable securities and agent's endowment: $\Cov(\CC,\EN_i)\cdot\Var^{-1}[\CC]$. The
demand for a particular security $C_j$, for $j\in\{1,...,k\}$, is
a decreasing function of its covariance with agent's
endowment, a fact that supports the use of the M-V criterion for
risk management purposes. Indeed, we expect that when $\Cov(C_j,\EN_i)$
is negative, agent $i$ is willing to take a long
position on $C_j$ as a partial hedging transaction. It is also important to note that only the intercept point (and not the slope) of the linear function $Z_i$ depends on agent $i$'s endowment (see also Demange and Laroque \cite{DemLar95} for analogous and more detailed discussion).
\end{remark}

In the sequel, we will need some further notation. Let $\AAA_{n\times
k}\subset\R^n\times\R^k$ denote the set of matrices that represent
allocations of the vector $\CC$. More precisely, element
$a_{ij}$, $i\in\{1,...,n\}$ and $j\in\{1,...,k\}$ of an allocation
$A\in\AAA_{n\times k}$, stands for the units of security $C_j$
that agent $i$ buys (negative $a_{ij}$ means short position) and
the zero-net supply of tradeable securities implies that
$\sum_{i=1}^na_{ij}=0$, for each $j\in\{1,...,k\}$. Also, for an
allocation matrix $A\in\AAA_{n\times k}$, $\aa_i$ denotes its
$i$-th row, for each $i\in\{1,...,n\}$.

When agents do not behave strategically, the risk-sharing achieved by trading $\CC$ is
given through a competitive equilibrium (which coincides with the
CAPM, see among others Magill and Quinzii \cite{MagQui96}).

\begin{definition}\label{def:equilibrium}
For a given vector of securities $\CC\in(\ltwo)^k$, the pair
$(\pp^*,A^*)\in\R^k\times\AAA_{n\times k}$ is called a competitive
price-allocation equilibrium of $\CC$ if $Z_i(\pp^*)=\aa^*_i$ for each
$i\in\{1,...,n\}$.
\end{definition}

Taking into account equation \eqref{eq:quadr
demand}, we get the uniqueness of competitive price-allocation equilibrium and its characterization. Below,
$\gamma$ stands for the aggregate risk-aversion coefficient,
i.e., $1/\gamma=:\sum_{i=1}^n1/\gi$, and $\lambda_i:=\gamma/\gamma_i$ denotes the relative risk tolerance of agent $i$. Also, following the standard notation, we define $\EN_{-i}:=\sum_{j\neq
i}\EN_j$ and $\lambda_{-i}:=1-\lambda_i$, i.e., $\EN_{-i}$ is the aggregate endowment of the rest
of the agents and $\lambda_{-i}$ their aggregate relative risk tolerance. 
\begin{proposition}[CAPM]\label{pro:equilibrium with quadratic}
The unique competitive price-allocation equilibrium $(\pp^*,A^*)\in\R^k\times\AAA_{n\times k}$ of a vector of securities $\CC$ is given by
\begin{equation}\label{eq:PEquadr}
    \pp^*=\EE[\CC]-2\gamma\Cov(\CC,\EN)\quad\text{and}\quad\aa_i^*=\Cov\left(\CC,C_i^o\right)\cdot\Var^{-1}[\CC],
\end{equation}
where $\aa_i^*$ is the $i$-th row of matrix $A^*$ and
\begin{equation}\label{eq: optimal contracts}
C_i^o:=\lambda_i\EN_{-i}-(1-\lambda_i)\EN_i, \quad\text{for each } i\in\{1,...,n\}.
\end{equation}
\end{proposition}
Proposition \ref{pro:equilibrium with quadratic}
states that the equilibrium prices do not depend on the variance-covariance
matrix of securities, but only on their expectations and their
covariances with the aggregate endowment (this covariance is
usually called the \textit{covariance value} of the securities).
Under no market power, the equilibrium price of a security
$C_j$ increases as its covariance with the aggregate endowment
decreases, a feature that indicates that higher demand of a
particular security implies higher equilibrium price.

\subsection{Optimal risk-sharing}\label{subsec:optimal risk sharing}

In the case where there are no exogenous constrains or additional
transaction costs, agents freely design
securities that \textit{Pareto-optimally} share their risk
exposures and complete the market. More precisely, we define the
set $\A:=\{\KK=(K_1,...,K_n)\in(\ltwo)^n:\sum_{i=1}^nK_i=0\}$, which
contains all the possible risk-sharing securities,
where $K_i$ stands for the payoff of the security that agent $i$ receives.$^{18}$\footnote{$^{18}$ 
The requirement that the sum of these securities is zero implies
that risk-sharing transaction is cleared out and there is no
further risk added to the market. In Section \ref{sec: noise trader}, we withdraw the assumption of zero-net supply, in order to examine how a non-constant  risk exposure submitted by noise traders affects the main results of the paper.} The formal definition of the
optimal risk-sharing follows.

\begin{definition}\label{def: Patero optimal}
A vector of securities $\KK^o\in\A$ is a Pareto-optimal
risk-sharing if for all other $\KK\in\A$ the following implication
holds:\\
If for some $i$, $\UU_i(\EN_i+K_i)>
    \UU_i(\EN_i+K^o_i)$, then $\exists j\neq i$ such that $\UU_j(\EN_j+K_j)<
    \UU_j(\EN_j+K^o_j).$
\end{definition}
Note that Pareto-optimal risk-sharing equilibrium assumes no market power, and all the agents submit their true risk exposures for the sharing. It is well-known that under M-V preferences, the optimal
risk-sharing securities are linear functions of the agents'
endowments (see among others Demange and Laroque \cite{DemLar95}, Wilson \cite{Wil68}). It follows
that we can restrict the problem of finding the optimal sharing to
finding the price-allocation of the agents' vector of endowments
$\EEE:=(\EN_1,...,\EN_n)\in(\ltwo)^n$. 
Proposition \ref{pro:equilibrium with quadratic} yields that the
competitive price-allocation equilibrium of $\EEE$ is the
pair $(\pp^o,A^o)\in\R^n\times\AAA_{n\times n}$, where
$\pp^o=\EE[\EEE]-2\gamma\mathbf{1}_n\cdot\Var[\EEE]$, 
$\mathbf{1}_n:=(1,...,1)$ and the elements of $A^o$ are $a^o_{ii}=-\lambda_{-i}$ and $a^o_{ij}=\lambda_i,$ for $j\neq i$.$^{19}$\footnote{$^{19}$The exponent $^o$ in the notation refers to the Pareto-optimal risk-sharing, i.e., when agents are allowed to trade their endowments and do not behave strategically, see also the securities defined in \eqref{eq: optimal contracts}. One the other hand, for equilibrium prices and allocations in the incomplete market-setting have notation with exponent $^*$.} The formal statement is given in the following proposition (its
proof is placed in Appendix \ref{sec:proofs}).

\begin{proposition}\label{pro:optimal sharing}
Let $(\pp^o,A^o)\in\R^n\times\AAA_{n\times n}$ be the 
competitive price-allocation equilibrium of the vector of securities $\EEE$.
Then, the vector of securities $A^o\cdot(\EEE -\pp^o)\in\A$ is the
unique Pareto-optimal risk-sharing.
\end{proposition}


At the optimal risk-sharing transaction, agent $i$
gets the total payoff $\aa^o_i\cdot\EEE-\aa^o_i\cdot\pp^o$, where
$\aa^o_i$ denotes the vector
$(a^o_{i1},...,a^o_{in})\in\mathbb{R}^n$. Note that $\aa^o_i\cdot\EEE=C_i^o,$ with $C_i^o$ defined in 
\eqref{eq: optimal contracts}. Hence, according to the
optimal risk-sharing rules, agent $i$ gets security $C_i^o$
(hereafter called \textit{optimal-sharing security}) and pays the
price
\begin{equation}\label{eq: optimal prices}
\pi_i^o:=\aa^o_i\cdot\pp^o =\EE[C_i^o]-2\gamma\Cov(C_i^o,\EN).
\end{equation}
In other words, under no use of market power, agent $i$
is going to short a part of her true random endowment and long
equal parts of the other agents' aggregate endowment.$^{20}$\footnote{$^{20}$From Proposition \ref{pro:optimal sharing} it also follows that the pair of the vector of securities $(C_i^o)^n_{i=1}$ and the pricing kernel $\pi^0:\ltwo\mapsto\R$, defined as $\pi^o[C]=\EE[C]-2\gamma\Cov(C,\EN)$, for $C\in\ltwo$, is in fact the unique Arrow-Debreu equilibrium among the $n$ agents.}

\subsection{Utility surpluses and the inefficiency measure}

One of the main goals of this paper is the comparison of each individual agent's utility surplus at different equilibria. For this, at any price $\pp$, we denote the utility of agent $i$ at her demand as 
\[v_i(\pp;\CC)=\UU_i(\EN_i+Z_i(\pp)\cdot\CC)-Z_i(\pp)\cdot\pp.\]
The difference $v_i(\pp;\CC)-\UU_i(\EN_i)$ simply measures the utility surplus that agent $i$ gets by buying her 
demand on $\CC$ at price $\pp$. The advantage of using $v_i(\pp;\CC)$ as a
measure of the agent's utility surplus from a transaction is that it is
measured in num\'{e}raire units and therefore can be used for
comparisons among different equilibria and agents. When the market setting is complete, we use the simplified notation $v_i^o:=v_i(\pp^o;\EEE)$. From Propositions \ref{pro:equilibrium with quadratic} and \ref{pro:optimal sharing}, we readily get the agent $i$'s utility at both market settings, provided that none of the agents applies any kind of strategic behavior.
\begin{corollary}\label{pro:quadratic equilibrium level}
The utility of agent $i$ at the Pareto-optimal risk-sharing is given by 
\begin{equation}\label{eq:level at optimal sharing}
v^o_i=\gi\Var[C^o_i]+\UU_i(\EN_i),
\end{equation}
while the utility at the competitive price-allocation equilibrium
of a security vector $\CC$ is 
\begin{equation}\label{eq:quadratic equilibrium level}
v_i(\pp^*;\CC)=\gi\aa_i^*\cdot\Var[\CC]\cdot\aa_i^*+\UU_i(\EN_i)=
\gi\Cov(\CC,C^o_i)\cdot\Var^{-1}[\CC]\cdot\Cov(\CC,C^o_i)+\UU_i(\EN_i).
\end{equation}
\end{corollary}

It follows that the utility
\textit{``loss''} of each agent when the risk-sharing is in
incomplete market and not in complete one, that is the difference
$v^o_i-v_i(\pp^*;\CC)$, is always non-negative and equal to zero if and only if the optimal security of agent $i$ belongs to the span of $\CC$. We state this fact in the
following proposition, the proof of which follows standard
arguments.

\begin{proposition}\label{pro:loss of utilities}
Each individual agent suffers a loss of utility from market
incompleteness. This loss is zero for agent $i$ if and only if
$C_i^o\in\spa(\CC)$.
\end{proposition}

\begin{remark}\label{rem: interprerteation of term}
The term $\Cov(\CC,C^o_i)\cdot\Var^{-1}[\CC]\cdot\Cov(\CC,C^o_i)$ in \eqref{eq:quadratic equilibrium level}
has a nice interpretation in the case where both $\CC$ and $\EEE$
follow multivariate normal distribution. Indeed, an application of the projection theorem (see among others, Chapter 1 of the book by Brunnermeier \cite{Bru01}) yields that
the conditional distribution of $C^0_i$ given $\CC$, remains normal with conditional variance given by
$\Var[C_i^o|\CC]=\Var[C_i^o]-\Cov(\CC,C^o_i)\cdot\Var^{-1}[\CC]\cdot\Cov(\CC,C^o_i).$
This implies the representation 
$v_i(\pp^*;\CC)=\gi\left(\Var[C_i^o]-\Var[C_i^o|\CC]\right)+\UU_i(\EN_i)$, which connects the utility surplus with the information on $C_i^o$ that is contained in $\CC$. Namely, the more information about $C_i^o$ is
revealed from $\CC$, the higher the agent $i$'s utility surplus is.
Note also that when $C_i^o$ is measurable with respect to the
information generated by $\CC$, $v_i(\pp^*;\CC)$ gets its maximum
value, and when $C_i^o$ and $\CC$ are independent, agent $i$ is
indifferent between trading and not trading $\CC$ at its equilibrium price.
\end{remark}

Regarding the aggregate utility, we point out that  at the optimal risk-sharing transaction 
$\sum_{i=1}^n\UU_i(\EN_i+C_i^o-\pi^o_i)=\EE[\EN]-\gamma\Var[\EN],$
and hence the difference
$\sum_{i=1}^n\UU_i(\EN_i+C_i^o-\pi^o_i)-\sum_{i=1}^n\UU_i(\EN_i)=
\sum_{i=1}^n\gi\Var[\EN_i]-\gamma\Var[\EN]$ is the
\textit{maximum aggregate utility surplus} that the market gets from the optimal risk-sharing.$^{21}$\footnote{$^{21}$Note that Pareto optimality and the M-V preferences guarantee
that
$\sup_{\KK\in\A}\sum_{i=1}^n\UU_i(\EE_i+K_i)=\sum_{i=1}^n\UU_i(\EE_i+C^o_i)$
(see also the proof of Proposition \ref{pro:optimal sharing}).} This allows us to consider the difference between the aggregate utilities in optimal
risk-sharing and in any realized (suboptimal) risk-sharing, as a measure of \textit{risk-sharing inefficiency}, i.e.
\begin{equation}\label{eq:definition inefficiency}
    \text{Risk-Sharing Inefficiency = Optimal Aggregate Utility
    }-\text{ Realized Aggregate Utility}.^{22}\footnote{$^{22}$This measure of aggregate
    loss of utility is also used in Acharya and Bisin \cite{AchaBin05}, while
    in Vayanos \cite{Vay99} the aggregate utilities are compared in terms of the certainty equivalent, which coincide with our comparison, under M-V preferences.}
\end{equation}

\bigskip

\section{Best Responses in Risk-Sharing Transactions}\label{sec:Agent's Strategic Behavior in Risk Sharing Market}

As emphasized in the introductory section, financial
risk-sharing transactions are normally among relatively few agents. Modelling of such \textit{oligopolies} should include
agents' best responses to other agents' actions. This section establishes a novel model for the agents' strategic
behavior in risk-sharing markets, which can be considered also as a (predatory trading) strategy implemented by an agent who has asymmetric information on other agents' endowments or demand functions. 

\subsection{Best response in complete risk-sharing markets}\label{subsec:best responses}

Proposition \ref{pro:optimal
sharing} states that the optimal way to share any vector of
submitted random endowments is through the (linear)
optimal sharing rules. Given that agents have agreed to adopt this transaction set-up, we ask whether it is preferable for an individual agent to submit 
her true random endowment or to report a different
risk exposure.

More precisely, let's consider agent $i$, and assume that she knows
the aggregate risk exposure of the rest of the agents. According
to optimal sharing rule, if she reports endowment $\EN_i$,
her position at the equilibrium will be $C_i^o-\pi^o_i$ (see
the payoffs in \eqref{eq: optimal contracts} and the prices in \eqref{eq: optimal prices}).
However, she may exploit the other agents' hedging needs that stem
from $\EN_{-i}$ and drive the security-designing to
more preferable security and price that the other agents are still willing
to offset. The fact that the optimal sharing rules are given linear functions of
the submitted endowments means that proposing securities and pricing kernel is in fact equivalent to submitting a corresponding endowment. Therefore, 
agent $i$ can achieve a desirable equilibrium by submitting an appropriate endowment in the sharing scheme. Following the terminology of game theory, we call an
optimal choice for submitted endowment \textit{best-endowment
response}. 

Intuitively, knowing endowment $\EN_{-i}$, agent $i$ has two goals: First to hedge as effectively as possible her true endowment (recall that all agents are assumed risk averse), and second to exploit the hedging needs of her counterparties and get a better cash compensation. Regarding the first goal, agent $i$ has to submit at least some part of her true endowment and achieve a partial hedging. On the other hand, she may declare exposure to the risk that the other agents possess, since by doing so, she increases the supply (and hence decreases the prices) of the securities she is going to buy.

More formally, given that her counterparties have reported aggregate
endowment $\EN_{-i}$, if agent $i$ reports as her random endowment
some random variable $B\in\ltwo$, she gets the security
with payoff 
\begin{equation}\label{eq:optimal contract when B is reported}
    C_i^o(B):=\lambda_i\EN_{-i}-\lambda_{-i} B,
\end{equation}
and the accumulated cash she has to pay is
\begin{equation}\label{eq:price paid when B is reported}
\pi_i^o(B):=\lambda_i\left(\EE[\EN_{-i}]-2\gamma\Cov(\EN_{-i},\EN_{-i}+B)\right)-\lambda_{-i}\left(\EE[B]-2\gamma\Cov(B,\EN_{-i}+B)\right).
\end{equation}
Hence, her utility after the transaction is
\begin{equation}\label{eq:utility when B is reported}
    G_i(B;\EN_{-i}):=\UU_i(\EN_i+C_i^o(B)-\pi_i^o(B))=\EE[\EN_i+C_i^o(B)]-\gamma_i\Var[\EN_i+C_i^o(B)]-\pi_i^o(B).
\end{equation}
Therefore, the best-endowment
response of agent $i$ is the solution of the following
maximization problem, where $\ltwo$ is the set of her strategic choices
\begin{equation}\label{eq:best response problem}
    \Br_i:=\underset{B\in\ltwo}{\argmax}\{G_i(B;\EN_{-i})\}.^{23}\footnote{$^{23}$Note that letting the set of
strategic choices be equal to $\ltwo(\Omega,\FF,\PP)$ implies that
agent's reported endowment is measurable with respect to the
information that is generated by the true endowments, i.e., all agents face the same $\sigma$-algebra.}
\end{equation}
It is important to point out again that $\Br_i$ (if it exists and is unique) directly determines the payoff of
the security that agent $i$ gets and the price she
pays at the equilibrium. In other words, submitting endowment $\Br_i$ for
sharing is equivalent to proposing the security $\Cr_i:=C^o_i(\Br_i)$,
determined by equation \eqref{eq:optimal contract when B is
reported}, and the induced cash compensation $\pir_i:=\pi_i^o(\Br_i)$
given in \eqref{eq:price paid when B is
reported}. The
solution of problem \eqref{eq:best response problem} is stated in
the following proposition.

\begin{proposition}\label{pro:best response}
For each $i\in\{1,...,n\}$, the unique (up to constants) best-endowment response of agent $i$, when the rest of the agents have
submitted aggregate endowment $\EN_{-i}$, is given by
\begin{equation}\label{eq:best response}
    \Br_i=\frac{1}{1+\lambda_i}\EN_i+\frac{\lambda_i^2}{1-\lambda_i^2}\EN_{-i}.
\end{equation}
\end{proposition}

By submitting endowment $\Br_i$, agent $i$ gets security with payoff $\Cr_i$ instead of
$C^o_i$, where we readily calculate that
$\Cr_i=C_i^o/(1+\lambda_i)$. Thus, according to best-endowment response, agent $i$
gets only a fraction of the Pareto-optimal risk-sharing
security, whereas the price she pays is less than
$\pi_i^o/(1+\lambda_i)$, and in fact
\[\pir_i=\frac{\pi_i^o}{1+\lambda_i}-2\gamma\frac{\lambda_i\Var[C_i^o]}{\lambda_{-i}(1+\lambda_i)^2}.\]
A number of further observations are worth some attention. First,
the best response $\Br_i$ is invariant on the probability
distribution of the random endowments.$^{24}$\footnote{$^{24}$This outcome is based on the common-beliefs assumption and the imposed M-V preferences. In particular, under common beliefs, M-V preferences imply that the optimal risk-sharing rules do not depend on the probability distribution of the endowments. This independence is endowed to the optimizer of problem \eqref{eq:best response problem}. In more general preferences' model however, risk-sharing rules, and hence the corresponding problem \eqref{eq:best response problem}, do depend on the probability distribution of the involved endowments (see for example Jouini et al.~\cite{JouSchTou08}). The generalization of the best-endowment response in such models is a very challenging problem and left for future research.} Also, given that agent $i$
knows the endowments of the rest of the agents, the best-endowment
response strategy indicates that she should share only a fraction
of her risk exposure and report exposure to the risk that the
other agents face. In this way, she increases (resp.~decreases) the
demand of the securities she sells (resp.~buys) at the risk-sharing
transaction, which results in a better total cash compensation, $\pir_i$. This strategy could well be considered as a predatory trading (in the spirit of Brunnermeier and Pedersen \cite{BruPed05}). Agent $i$ exploits the hedging needs of her counterparties and acts strategically aiming not only to a good hedging, but also to a higher price of the securities she sells. 

The agent's utility after reporting the best-endowment
response is
$G_i(\Br_i;\EN_{-i})=\gi\Var\left[C_i^o\right]/(1-\lambda_i^2)+\UU_i(\EN_i),$
which means that the surplus of utility of agent $i$ from applying
this strategic behavior is equal to
$\gi\Var[C_i^o]\lambda_i^2/(1-\lambda_i^2)$ or in other words,
compared to utility surplus in \eqref{eq:level at optimal sharing}, the percentage
increase of utility is $\lambda_i^2/(1-\lambda_i^2)$ (an increasing function of $\lambda_i$).
Therefore, the application of such predatory trading is more beneficial
for agents with relatively low risk aversion; while for homogeneous
agents, the percentage increase of utility equals to
$1/(n^2-1)$. 

Note also that the difference between
$\Br_i$ becomes equal (up to constants) to $\EN_i$ if and only if $C_i^o$ is a constant, which essentially means that agent $i$ has nothing to trade with the rest of the agents. Furthermore, and as expected, the difference between
$\Br_i$ and $\EN_i$ reduces as the number of participating agents increases (the larger the market becomes, the less effective the use of market power is).
\begin{remark}\label{rem:speculator complete}
The case of a speculator, i.e., an agent
with constant endowment, is of special interest. In the optimal risk-sharing equilibrium, a speculator gets $\lambda_i$ units of the other agents' aggregate endowment, in exchange of a risk premium equal to $\lambda_i(2\gamma\Var[\EN_{-i}]-\EE[\EN_{-i}])$. According to best-endowment response however, she exploits her knowledge on the other agents' endowment and submits risk exposure to $\EN_{-i}$ too. In this way, she increases the supply of $\EN_{-i}$, which eventually yields to a higher risk premium for those who buy $\EN_{-i}$. At resulting equilibrium, not only she gets less units of $\EN_{-i}$ (which means less risk for her), but also a higher premium. 
\end{remark}

\subsection{Best response in incomplete risk-sharing markets}\label{subsec:best response constrained}

In the case of incomplete market-setting, where agents share their risk
only through the transaction of a given vector of securities, each agent $i$ reports her demand function $Z_i(\pp)$ and the (competitive) equilibrium is reached at 
the price vector that sums
the demands to zero. In view of the best-endowment response,
we conjecture that agents have motive to declare demand functions that do not reflect their true hedging needs. Indeed, it is shown below that when an agent does share some risk with the other agents, it is never optimal to submit her true demand function.

The study of the strategic behaviors under incomplete markets through the submission
of demand functions goes back at least to Kyle \cite{Kyle89} (see also Rostek and Weretka \cite{RosWer15} and the reference therein).
Following this literature, we consider agent $i$ and
suppose that she knows (or can exact) the aggregate demand
function of the rest of the agents. Recall from equation
\eqref{eq:quadr demand} that under M-V
preferences, each agent's demand function is linear with slope
equal to $-\Var^{-1}[\CC]/2\gamma_i$. This implies that the aggregate demand
function of the other $(n-1)$ agents is also linear and given
by  $Z_{-i}(\pp):=\sum_{j\neq
i}Z_j(\pp)=\left[(\EE[\CC]-\pp)/2\gamma_{-i}-\Cov(\CC,\EN_{-i})\right]\cdot\Var^{-1}[\CC]$,
where we use the notation $\gamma_{-i}:=(\sum_{j\neq
i}1/\gamma_{j})^{-1}$. Then,
agent $i$ submits a demand function that clears out the
market at the price that maximizes her own utility. In contrast to the related literature$^{25}$\footnote{$^{25}$For instance, the game developed in Rostek and Weretka \cite{RosWer15} or Weretka \cite{Wer11} does not allow agents to submit different hedging needs than those induced by their true endowments. The key difference is that our game is played on the intercept point of the demand functions not on their slopes.}, we do not assume that agent $i$ responds to the slope of demand $Z_{-i}(\pp)$ (which does not contain the agents' endowments), but rather we ask: \textit{Which is the best equilibrium price
for agent $i$, given the aggregate demand of the other agents?}$^{26}$\footnote{$^{26}$As explained below, this is a more appropriate way of modelling strategic behavior under endowments that do not belong to the span of the tradeable securities, since it allows us to see how the best responses reflect different hedging needs than the true ones.} 

Hence, agent $i$'s 
problem is written as
\begin{equation}\label{eq:utility given Zi}
    \ppr_i:=\underset{\pp\in\R^k}{\argmax}\left\{\phi_i(\pp;Z_{-i}(\pp))\right\},
\end{equation}
where $\phi_i(\pp;\aa):=\UU_i(\EN_i-\aa\cdot\CC+\aa\cdot\pp)$, for
$\pp,\aa\in\R^k$. Below, we state how the competitive price-allocation 
equilibrium changes when an agent's market power
stems from the asymmetric information on the other agents'
demands.
\begin{proposition}\label{pro:best price response}
The price-allocation equilibrium of a vector of securities $\CC\in(\ltwo)^k$ that maximizes the utility of agent
$i$, for each $i\in\{1,...,n\}$, given the other agents' aggregate demand function, is provided by
\begin{equation}\label{eq:best price}
    \ppr_i= \EE[\CC]-2\gamma\Cov\left(\CC,\frac{\EN_i}{1+\lambda_i}+\frac{\EN_{-i}}{1-\lambda_i^2}\right),
\end{equation}
\begin{equation}\label{eq:best allocation}
\aar_i=\frac{\aa_i^*}{1+\lambda_i}\quad\text{and for all }j\neq i,\quad\aar_j=\aa_j^*+\frac{\lambda_i\lambda_j}{1+\lambda_i}\Cov\left(\CC,\frac{\lambda_i}{\lambda_{-i}}\EN_{-i}-\EN_i\right)\cdot\Var^{-1}[\CC].
\end{equation}
\end{proposition}
It follows from \eqref{eq:best allocation} that the units of $\CC$ that agent $i$ gets after best-response strategy is reduced when compared to the competitive allocation. In addition, the \textit{effective} aggregate endowment in the
covariance part of the CAPM is equal to
$\EN_i/(1+\lambda_i)+\EN_{-i}/(1-\lambda_i^2)$
instead of $\EN$. In the
effective aggregate endowment, the volume on $\EN_i$ is reduced by
a percentage equal to $\lambda_i/(1+\lambda_i)$, whereas the
volume on $\EN_{-i}$ is increased by a percentage
equal to $\lambda_i^2/(1-\lambda_i^2)$. 
It is implied therefore that
agent $i$ could drive the market to price $\ppr_i$ if her
submitted demand reflects less exposure to her true endowment and
some exposure to the other agents' aggregate endowment (similarly to the best-endowment response). The exact demand
function of agent $i$ that clears out the market at price
$\ppr_i$ is called \textit{best-demand response}.

To formulate this best-response problem, we have to rewrite problem
\eqref{eq:utility given Zi} in terms of demand functions. For
this, having security vector $\CC$ fixed, we denote by $\ZZ_i$ the set of all demand functions that an
agent with M-V preferences and risk aversion $\gamma_i$ could submit. Since, all these demand functions
have the specific form of \eqref{eq:quadr demand}, we can
parametrize $\ZZ_i$ by the covariance vectors: A function
$z_i:\R^k \rightarrow\R^k$ belongs to $\ZZ_i$ if and only if there
exists a vector $\mathbf{c}\in\R^k$, such that
\begin{equation}\label{eq: demands in Z}
z_i(\pp)=\left(\frac{\EE[\CC]-\pp}{2\gi}-\mathbf{c}\right)\cdot\Var^{-1}[\CC].
\end{equation}
Note that vector $\mathbf{c}$ reflects the hedging needs, since it takes the position of the covariance between the agent's endowment and the tradeable securities. 
Taking the market-clearing into account, the best response of agent $i$ is to report the demand 
$\Zr_i\in\ZZ_i$, such that $\Zr_i(\ppr_i)+Z_{-i}(\ppr_i)=\mathbf{0},$ or equivalently to choose the
vector $\mathbf{c}^{\mathsf{r}}_i$, such that
$\left[(\EE[\CC]-\ppr_i)/2\gamma-\left(\Cov(\CC,\EN_{-i})+\mathbf{c}^{\mathsf{r}}_i\right)\right]\cdot\Var^{-1}[\CC]=\mathbf{0}.$
It follows that
$\Zr_i(\pp)=\left[(\EE[\CC]-\pp)/2\gi-\Cov(\CC,\Br_i)\right]\cdot\Var^{-1}[\CC],$
where the random variable $\Br_i$ is the best-endowment response
given in \eqref{eq:best response}. We summarize this discussion
in the following proposition, which states the clear connection
between the agent's best responses in complete and incomplete
market settings.
\begin{proposition}\label{pro: best demand and endowment}
For every vector of securities $\CC\in (\ltwo)^k$:
\begin{itemize}
\item[(i)] The best-demand response of an agent $i$ coincides with the demand
function that is associated with her best-endowment response.
\item[(ii)] The best-demand response equals to the true demand if and only if $\aa_i^*=\mathbf{0}$.
\item[(iii)] The difference of utilities at the best-response and at competitive equilibrium is
\begin{equation*}
\UU_i  \left(\EN_i+\aar_i\cdot(\CC-\ppr_i)\right) -\UU_i\left(\EN_i+\aa^*_i\cdot(\CC-\pp^*)\right)=\gi\frac{\lambda^2_{i}}{1-\lambda^2_i}\aa^*_i\cdot\Var[\CC]\cdot \aa^*_i.
\end{equation*}
\end{itemize}
\end{proposition}
\begin{remark}\label{rem:gains comparison in incomplete}
In view of Corollary \ref{pro:quadratic equilibrium level}, we have the following relation between the agent $i$'s utility surpluses \[\text{Utility surplus from best-response}=\frac{\lambda^2_{i}}{1-\lambda^2_i}\times\text{Utility surplus in competitive equilibrium.}\]
We conclude that the best-response strategy is more profitable when agent is relatively low risk averse and when her utility surplus from the competitive equilibrium (i.e., without strategic behavior) is large. 
\end{remark}

Proposition \ref{pro:best price response} states that although the strategy of best-demand response changes the equilibrium prices and reduces the size of the agent's positions, it does not change the direction of her positions. In particular, by submitting demand $\Zr_i$, agent $i$ increases
(resp.~decreases) the demand of the securities she is going to short
(resp.~long) at the equilibrium, and thus creates a beneficial impact on the
equilibrium prices, i.e., a predatory trading strategy. A measure of this price impact can simply be
given by the difference $\ppr_i-\pp^*$. For any vector $\CC$, we readily calculate that this difference equals to
$-2\gamma\lambda_{i}\Cov(\CC,C_i^o)/(1-\lambda^2_{i})$. In view of Proposition \ref{pro:equilibrium with quadratic}, each element of this vector is positive (resp.~negative) for the securities that agent $i$ sells (resp.~buys).

\begin{remark}\label{rem:speculator}
The case of a speculator is distinctive (connect it with the discussion in Remark \ref{rem:speculator complete}). In the competitive equilibrium, a speculator satisfies the hedging needs of the other agents by buying (resp.~selling) the securities they want to sell (resp.~buy). Under this perspective, a speculator could be considered as a (risk averse) market maker. For example, if the payoff of the security, say $C_j$, is negatively correlated with endowment $\EN_{-i}$, the speculator is supposed to take short position on $C_j$. According to her best-demand response however, knowing the hedging needs of her counterparties, she is going to submit a demand function that reveals some hedging needs (which are in fact of the same direction as the other agents' aggregate demand). She does so in order to increase the demand on $C_j$ and eventually increases its market value. This is a clear predatory trading. Note that at equilibrium she still takes a short position on $C_j$ (as allocation \eqref{eq:best allocation} indicates), but her predatory trading strategy increases the risk premium that she is paid by the other agents.
\end{remark}
\bigskip

\section{Nash Risk-Sharing Equilibria}\label{sec:nash equilibria}

The existence and the uniqueness of best-responses allow us to take the next step and examine whether and
how these thin markets equilibrate. Each agent responds to
other agents' choices forming a type of pure-Nash game, where (depending on the market completion) the
strategic sets of choices are the set of endowments or demands.

In this section, we study the existence and the uniqueness of the  Nash equilibria, and ask whether the agents' motive to  declare different risk exposures than their true ones still holds, when all agents act strategically. We also ask whether there are agents that
benefit from the games in thin risk-sharing markets and get higher utility surpluses.

\subsection{Nash equilibrium in complete risk-sharing market}\label{subsec:Nash complete}

The way that agents use their market power in the complete
risk-sharing market is provided by optimization problem \eqref{eq:best
response problem}. Namely, each agent declares the random
endowment she chooses to share (the one indicated by
solution \eqref{eq:best response}) or equivalently proposes securities that
are in line with the optimal sharing rules (see payoff in
\eqref{eq:optimal contract when B is reported} and prices in \eqref{eq:price
paid when B is reported}). This procedure sets the conditions of
the Nash game in the complete risk-sharing market, and the
equilibrium is reached at the security payoffs that all the agents
agree on. We call this equilibrium \textit{Nash risk-sharing
equilibrium}.$^{27}$\footnote{$^{27}$This is a pure-Nash
equilibrium where the set of strategic choices is in fact $\ltwo$.
The definition presupposes that the additional costs and the time
of the negotiation are negligible, both of which are standard assumptions
in the literature of the Nash equilibria.} Before we give the exact definition, we recall from
\eqref{eq:utility when B is reported} that the utility of
agent $i$, when she submits endowment $B$ and the other
agents have submitted endowment $B_{-i}:=\sum_{j\neq
i}B_j$ is written as
\begin{equation}\label{eq: Gi in Nash}
G_i(B;B_{-i})=\UU_i(\EN_i+C_i^o(B)-\pi^o_i(B))=\EE[\EN_i+C_i^o(B)]-\gamma_i\Var[\EN_i+C_i^o(B)]-\pi^o_i(B).
\end{equation}
\begin{definition}\label{def: Nash equilibrium}
We call a vector of random variables $(
\Bn_1,...,\Bn_n)\in(\ltwo)^n$ Nash equilibrium endowments if for
each $i\in\{1,...,n\}$
\begin{equation}\label{eq:condition for Nash equilibrium}
    G_i(\Bn_i;\Bn_{-i})\geq G_i(B;\Bn_{-i}),\quad\quad\text{for all } B\in\ltwo.
\end{equation}
The induced risk-sharing securities $(\Cn_1,...,\Cn_n)$ given by
\begin{equation}\label{eq: Nash contracts def}
    \Cn_i:=C^o_i(\Bn_i)=\lambda_i\Bn_{-i}+\lambda_{-i}\Bn_i,\quad\quad\text{for each }i\in\{1,...,n\},
\end{equation}
are called Nash risk-sharing securities.$^{28}$\footnote{$^{28}$Throughout the paper, exponent $^{\diamond}$ refers to Nash equilibrium features.}
\end{definition}

If a vector of Nash equilibrium endowments exists, the prices of
the induced risk-sharing securities $(\Cn_1,...,\Cn_n)$ are
determined by the pricing rule \eqref{eq: optimal prices}, where
the considered aggregate endowment is the Nash equilibrium one,
hereafter denoted by $\BBn:=\sum_{i=1}^n\Bn_i$. Hence, for each
$i\in\{1,...,n\}$, the Nash equilibrium price of security $\Cn_i$
is $\pin_i:=\pi_i^o(\Bn_i)=\EE[\Cn_i]-2\gamma\Cov(\Cn_i,\BBn)$. The following theorem gives the explicit form of the
\textit{unique} Nash risk-sharing equilibrium.
\begin{theorem}\label{thm:nash}
There exists a unique (up to constants) Nash risk-sharing
equilibrium characterized as follows: For each $i\in\{1,...,n\}$,
  the Nash equilibrium endowment of agent $i$ is
 \begin{equation}\label{eq:B*i}
    \Bn_i=\lambda_{-i}\EN_i+\lambda_i^2\BBn, 
\end{equation}
where
\begin{equation}\label{eq:B*}
    \BBn=\frac{1}{\alpha}\sum_{j=1}^n\lambda_{-j}\EN_j
\end{equation}
and the constant $\alpha\in\R_+$ is defined by
$\alpha:=1-\sum_{j=1}^n\lambda_j^2$. The Nash risk-sharing security that agent $i$ gets has payoff
     \begin{equation}\label{eq:C*}
    \Cn_i=\lambda_i\frac{\lambda_{-i}}{\alpha}\sum_{j\neq i}\lambda_{-j}\EN_j-\lambda_{-i}\left(1-\frac{\lambda_i\lambda_{-i}}{\alpha}\right)\EN_i,
    \end{equation}
and the utility of agent $i$ at the Nash
equilibrium is
\begin{equation}\label{eq:utility gain nash}
v^o_i(\pin_i):=\UU_i(\EN_i+\Cn_i-\pin_i)=\gamma_i\frac{1+\lambda_i}{1-\lambda_i}\Var[\Cn_i]+\UU_i(\EN_i)=\gamma_i\frac{\gamma_i+\gamma}{\gamma_i-\gamma}\Var[\Cn_i]+\UU_i(\EN_i).
\end{equation}
\end{theorem}
 
A number of observations are worth some emphasis. First, in consistence with Proposition \ref{pro:best response}, in Nash equilibrium each agent submits only a fraction of her true endowment and reports exposure to the endowments of the other agents. Indeed, the coefficient of $\EN_i$ in the equilibrium endowment $\Bn_i$ equals to $\lambda_{-i}+\lambda_i^2\lambda_{-i}/\alpha$, which is strictly less than 1 for each $i$. As a result, the risk-sharing transaction is significantly departed from the Pareto-optimality. Not only the security that
each agent gets at equilibrium is different than the optimal risk-sharing
one, but also, in general, the aggregate shared endowment
deviates from the true one (they coincide only when agents are
homogeneous, see Corollary \ref{cor:sharing game} below). In particular, one can readily check that
$0<1-\lambda_i\lambda_{-i}/\alpha<1,$ which, in view of \eqref{eq:C*}, means that
at the Nash equilibrium each agent still sells some of her true
endowment, but the size is \textit{always} smaller than in Pareto-optimal transaction (recall from \eqref{eq: optimal contracts} the payoff of security $C_i^o$). Also, for each agent $i$, the quantity of her true risk exposure that she sells at Nash equilibrium is an increasing function of her risk-aversion coefficient (i.e., decreasing function of $\lambda_i$), meaning that lower risk averse agents apply the strategic behavior more intensely.

On the other hand, the quantity of $\EN_j$ that agent $i$ buys at Nash equilibrium is not necessarily less than the corresponding quantity in the competitive equilibrium. For this, we compare the coefficients $\lambda_i\lambda_{-i}\lambda_{-j}/\alpha$ and $\lambda_i$ (recall again \eqref{eq: optimal contracts}) and observe that only in the special case of two agents, the size in Nash equilibrium is always less than the Pareto-optimal one (see also the indicative Table \ref{table} below). For $n\geq 3$, the inequality $\lambda_{-i}\lambda_{-j}/\alpha<1$ holds if $\lambda_i$ and $\lambda_j$ are sufficiently large$^{29}$\footnote{$^{29}$It is a matter of simple calculations to get an equivalent inequality of condition $\lambda_{-i}\lambda_{-j}/\alpha<1$. Indeed, for $n=3$, $i=1$ and $j=2$, the latter inequality is equivalent to
$$\frac{\sum_{\zeta=3}^n\lambda^2_{\zeta}}{\sum_{\zeta=3}^n\lambda_{\zeta}}<(\lambda_1+\lambda_2)+\frac{\lambda_1\lambda_2}{\sum_{\zeta=3}^n\lambda_{\zeta}}.$$}, meaning that agent $i$ buys less of endowment $\EN_j$ if both agents $i$ and $j$ are sufficiently low risk averse. The latter implies that for relatively low risk averse agents, only a percentage of their true endowments is shared in Nash equilibrium. We summarize the main conclusions of this discussion below.
\begin{corollary}\label{cor:comparison} At Nash risk-sharing equilibrium:
\begin{itemize}
\item[(i)] Each agent reports only a fraction of her true endowment and some exposure to the endowments of the rest of the agents.
\item[(ii)] The total quantity of endowment $\EN_i$ that is shared in Nash equilibrium is less or equal to 1, if and only if
\begin{equation}\label{eq:comparison weights}
\frac{\sum_{j\neq i}\lambda_j^2}{\sum_{j\neq i}\lambda_j}\leq\lambda_i.
\end{equation}
\end{itemize}

\end{corollary}
Note that inequality \eqref{eq:comparison weights} holds strictly when $n=2$, while for $n\geq 3$ the equality holds when agents have the same risk aversion.
\begin{remark}\label{rem:asymptotics}
Another aspect that highlights the importance of the agents' risk
aversion is the asymptotic interaction of agents' behaviors.
Namely, we observe that
$\lzer$-$\underset{\gi\rightarrow\infty}{\lim}\Bn_i=\EN_i$, for every
$\gamma_{-i}\in\mathbb{R}_+$; while
$\lzer$-$\underset{\gamma_{-i}\rightarrow 0}{\lim}\Bn_i=\EN_i$, for every
$\gi\in\mathbb{R}_+$. Intuitively the latter limit implies that agents with risk preferences
close to risk neutrality tend to \textit{dominate} the Nash risk-sharing
game. 
\end{remark}

The fact that the aggregate shared
risk is different than $\EN$ implies (as expected) inefficient sharing of risks. Thanks to the formulas of the equilibrium securities, we
are able to measure this inefficiency explicitly (recall the inefficiency measure defined in \eqref{eq:definition inefficiency}).
\begin{corollary}\label{cor:inefficiency}
The risk-sharing inefficiency caused by the agents' use of market
power is given by
\begin{equation}\label{eq:loss}
\sum_{i=1}^n\UU_i(\EN_i+C_i^o-\pi_i^o)-\sum_{i=1}^n\UU_i(\EN_i+\Cn_i-\pin_i)=\sum_{i=1}^n\gamma_i\Var[\EN_i-\Bn_i]-\gamma\Var[\EN-\BBn].
\end{equation}
The inefficiency vanishes if and only if the
Pareto-optimal securities are constants, i.e., there is no mutually beneficial transfer of risk among the agents.
\end{corollary}
Hence, even when agents have agreed to apply the optimal sharing rules to submitted endowments, strategic behavior always yields loss of efficiency in any case where some risk is to be transferred. No risk-sharing among agents means that Pareto-optimal securities are constants, which holds if and only if $\gamma_i\EN_i=\gamma_j\EN_j$
(modulo constants), for each $i,j\in\{1,...,n\}$. 

Intuitively, the inefficiency of the risk-sharing should decrease to zero as
the number of agents increases, meaning that the market power of
each individual agent is getting negligible. Upon some mild
assumptions, the following proposition verifies this standardized fact.

\begin{proposition}\label{pro:n goes to infty}
Consider a sequence of agents with random endowments
$(\EN_i)_{i\in\N}$ and risk-aversion coefficients
$(\gamma_i)_{i\in\N}$. If $(\EN_i)_{i\in\N}$ is uniformly bounded in $\ltwo$ and there exist positive constants $c_l$
and $c_u$ such that $c_l\leq
\gamma_i\leq c_u$ for each $i\in\N$, the risk-sharing inefficiency
goes to zero as $n\rightarrow\infty$.
\end{proposition}
It should also be pointed out that the enlargement of the
market increases the efficiency, even if the new agents are
speculators (that is, they do not have any risk exposure to share).
In fact, although agents with relatively low risk aversion tend to dominate thin markets, the total efficiency is increased,
even when the extra agents are low risk averse or/and without any
hedging needs.

\subsubsection{The case of two agents}
In order to give some further economic insights of Theorem \ref{thm:nash} and examine the utility surplus of each agent at Nash equilibrium, we
pay extra attention to the simplified case of two agents, which is summarized in
the Table \ref{table}. The main messages, however, of the following discussion can be generalized when $n$ is larger than two.

\begin{table}[!htbp]
  \begin{center}
\begin{tabular}{|c|c|c|}
  \hline
   & \underline{Pareto Sharing} & \underline{Nash Sharing}  \\
\hline
  Aggregate submitted endowment & $\EN$ & $\BBn=\frac{\gamma_1\EN_1+\gamma_2\EN_{2}}{2\gamma}$ \\
  \hline
  Submitted endowment of agent 1 & $\EN_1$ & $\Bn_1=\frac{2\gamma_1+\gamma_2}{2(\gamma_1+\gamma_2)}\EN_1+\frac{\gamma_2^2}{2\gamma_1(\gamma_1+\gamma_2)}\EN_2$ \\
  \hline
  Purchased security by agent 1 & $C_1^o=\frac{\gamma_{2}\EN_{2}-\gamma_1\EN_1}{\gamma_1+\gamma_{2}}$ & $\Cn_1=\frac{C_1^o}{2}$ \\
\hline
  Utility surplus of agent 1 & $\gamma_1\Var\left[\frac{\gamma_1\EN_1-\gamma_2\EN_2}{\gamma_1+\gamma_2}\right]$ & $\frac{\gamma_1+2\gamma_2}{4}\Var\left[\frac{\gamma_1\EN_1-\gamma_2\EN_2}{\gamma_1+\gamma_2}\right]$ \\
\hline  Inefficiency & 0 & $\frac{1}{\gamma_1+\gamma_2}\Var\left[\frac{\gamma_1\EN_1-\gamma_2\EN_2}{2}\right]$ \\
  \hline
\end{tabular}
  \end{center}
  \caption{{\footnotesize Comparison of Pareto and Nash risk-sharing equilibria when $n=2$.}}\label{table}
\end{table}

In the two-agent game, the Nash risk-sharing security is exactly half of the Pareto-optimal one, independently of agents'
risk-aversion coefficients. Furthermore, although the aggregate
utility is lower in Nash equilibrium, there exist
situations where some of the agents enjoy higher utility
when the game is played. We readily calculate that cash compensation in Nash equilibrium, $\pin_i=\EE[\Cn_i]-2\gamma\Cov(\Cn_i,\BBn)$, is always in favor
of the agent with lower risk aversion (when compared to the competitive one given in \eqref{eq: optimal prices}).
\begin{proposition}\label{pro:more gain in Nash}
Let $n=2$ and assume that $\gamma_1<\gamma_2$. Besides that
$\Cn_1=C_1^o/2$, when the price of $C_1^o$ is positive (resp.~negative)
the cash compensation $\pin_1$ is lower (resp.~higher) than $\pi_1^o/2$. Furthermore, agent 1 gets higher utility surplus at Nash equilibrium when compared to the optimal risk-sharing equilibrium if and only if $\gamma_1<2\gamma_{2}/3$. 
\end{proposition}
Therefore, market power and the induced Nash equilibrium have two
effects for agent 1: One negative, since she carries some of the
risk-sharing inefficiency, and one positive which stems from the
price impact. The total outcome is positive when her risk aversion is sufficiently lower than the one of her counterparty
(in particular when $\gamma_1<2\gamma_{2}/3$ or $\lambda_1>0.6$). We may conclude that
\textit{agents with sufficiently lower risk aversion benefit from
the oligopolistic structure of a risk-sharing transaction}. On the
other hand, an agent with higher risk aversion not only carries
some of the risk-sharing inefficiency, but also suffers from the
induced price impact (heavily when her counterparty's preferences
are close to risk neutrality).$^{30}$\footnote{$^{30}$A comparison
of the utility surpluses has also been studied in Palomino \cite{Pal96}, where it
is shown that there are cases where noise traders benefit from the non-competitive (Nash)
equilibrium prices.} 

This situation is connected to the aforementioned predatory trading strategy. In particular, it is shown that sufficiently lower risk averse agent could be considered as  predatory trader, even when her counterparty acts strategically. We recall that a predatory trader has two goals; one is to hedge her true endowment and the other to achieve a better cash compensation by exploiting the hedging needs of the other agent. When both agents apply this strategy, the aggregate utility surplus is reduced, whereas the individual utility surpluses depend on the risk aversions. As already pointed out, lower risk aversion means intenser implementation of the strategy, or in other words, higher risk tolerance means that the goal for better prices prevails the goal for better hedging. In the equilibrium, only the agent with sufficiently lower risk aversion benefits from predatory trading and in a sense exploits the stark hedging needs of her counterparty.

\begin{remark}\label{rem:comment on strategic set} 
The fact that low risk averse agents get more utility at Nash equilibrium stems from the agents' possibility to submit exposure on endowment they do not actually possess. When the set of strategic choices is restricted and does not allow this potentiality, the comparison of utility surpluses at the equilibria may not give the same outcome. An indicative example of a game under restricted set of strategic choices is developed in Appendix \ref{sec:best response with percentage}. Therein, agents are not allowed to submit any endowment, but rather they strategically choose only the size of their true endowments that they are going to submit for sharing. Under such restriction, it is shown that low risk aversion does not guarantee more utility surplus at Nash equilibrium.    
\end{remark}

\subsubsection{The case of homogeneous agents}
It is apparent from the above analysis that agents' heterogeneity in risk
preferences is an important factor for the outcome of the game. The large
majority of literature on Nash equilibria in risk-sharing
transactions assumes equal agents' risk aversions (see
for example Rahi and Zigrand \cite{RahZig09}, Rostek and Weretka \cite{RosWer15} and Vayanos \cite{Vay99}), which is only a
special case in the present set-up. Agents' homogeneity is
in fact the only non-trivial case where the aggregate Nash equilibrium  endowment is equal to the true one.
\begin{corollary}\label{cor:sharing game}
Assume that there exists a pair
$i,j\in\{1,...,n\}$, such that $\gamma_i\EN_i-\gamma_j\EN_j$ is not a constant. Then, $\BBn=\EN$ if and only
if agents are homogenous with respect to their risk aversions.
\end{corollary}
Equal agents' risk aversions also yield that the Nash risk-sharing securities are fractions
of the optimal ones: $\Cn_i=(n-1)C_i^o/n$, for each
$i\in\{1,...,n\}$. This
implies that the risk-sharing inefficiency equals to $(1/n^2)\left(\sum_{i=1}^n\Var[\EN_i]-\Var\left[\EN\right]/n\right)$, and  
is \textit{equally} shared among agents (see the inefficiency measure \eqref{eq:loss}).

\subsection{Nash equilibrium in incomplete risk-sharing market}\label{subsec:Nash incomplete}

For the risk-sharing equilibrium in the incomplete-market setting, we follow the arguments of subsection
\ref{subsec:best response constrained} and let the set of
strategic choices be the demand functions on the given vector
of securities, i.e., the sets $\ZZ_i$ for each $i\in\{1,...,n\}$.
We suppose that each agent responds to the others'
aggregate demand, aiming to drive the market to her preferable
price vector (recall problem \eqref{eq:utility given
Zi}). Naturally, the (Nash) equilibrium is the point at which the
submitted demand functions sum up to zero at a price which is
preferable for all agents. We define
$\ZZ:=\ZZ_1\times...\times\ZZ_n$ and give the exact definition
below.
\begin{definition}\label{def: Nash equilibrium of C}
A pair $(\ppn,\mathbf{\Zn})\in\R^k\times\ZZ$ is called
Nash price-demand equilibrium of a vector of securities
$\CC\in(\ltwo)^k$ if the following two conditions hold:
\begin{itemize}
    \item [(i)] $\sum_{i=1}^n\Zn_i(\ppn)=0$.
    \item [(ii)] For each $i\in\{1,...,n\}$, $\phi_i(\ppn;\Zn_{-i}(\ppn))\geq\phi_i(\pp;\Zn_{-i}(\pp))$, for all $\pp\in\R^k$.
\end{itemize}
For every Nash price $\ppn$, the vector
$(\Zn_1(\ppn),...,\Zn_n(\ppn))\in\AAA_{n\times k}$ is called Nash
allocation of $\CC$.
\end{definition}

In principle, this equilibrium can be seen as a \textit{variation of the CAPM}
in markets where one of the main assumptions of this model, that
all participants in the market are price-takers, is withdrawn.
Although games on demand functions generally equilibrate in a
continuum of Nash equilibria, in case of M-V preferences the Nash
equilibrium turns out to be unique.

\begin{theorem}[CAPM in Oligopoly Markets]\label{thm:Nash equilibrium price}
For every vector of securities $\CC\in(\ltwo)^k$, there is a
unique Nash price-demand equilibrium
$(\ppn,\mathbf{\Zn})\in\R^k\times\ZZ$ given by
\begin{equation}\label{eq:Nash equilibrium price of C}
    \ppn=\EE[\CC]-2\gamma\Cov(\CC,\BBn),
\end{equation}
and for each $i\in\{1,...,n\}$
\begin{equation}\label{eq:Nash demands}
        \Zn_i(\pp)=\left(\frac{\EE[\CC]-\pp}{2\gi}-\Cov(\CC,\Bn_i)\right)\cdot\Var^{-1}[\CC],
\end{equation}
where $\Bn_i$ and $\BBn$ are given in \eqref{eq:B*i} and
\eqref{eq:B*}. The induced Nash allocation is
\begin{equation}\label{eq: Nash equilbirium allocation}
    \Zn_i(\ppn)=\Cov(\CC,\Cn_i)\cdot\Var^{-1}[\CC],
\end{equation}
where $\Cn_i$ is given in \eqref{eq: Nash contracts def}, while the utility of agent $i$ at Nash equilibrium is 
\begin{equation}\label{eq: Nash level C}
    v_i(\ppn;\CC)=\UU_{i}(\EN_i)+\gamma_i\left(\frac{\gi+\gamma}{\gi-\gamma}\right)\Cov(\CC,\Cn_i)\cdot\Var^{-1}[\CC]\cdot\Cov(\CC,\Cn_i).
\end{equation}
\end{theorem}

It is important to note that the Nash equilibrium is valid even in the two-agent transaction. This comes in sharp contrast to the non-competitive equilibrium model developed in Kyle \cite{Kyle89} (see also Rostek and Weretka \cite{RosWer12}), where the equilibrium with two strategic players does not exist. The main reason for that is the imposed agents' strategic sets, which induce a game on the intercept points of agents' demand functions and give prominence to the importance of the agents' endowments. The two-agent equilibria are of special interest, since many of the real-world risk-sharing transactions (at least in primary level) is between only two institutions.  

Theorem \ref{thm:Nash equilibrium price} establishes a clear connection
between the Nash equilibria in complete and incomplete settings, arguing that in any risk-sharing 
transaction (complete or incomplete) that allows strategic behavior, the
considered agents' endowments should be $\Bn_i$ rather than
$\EN_i$. Indeed, Nash equilibrium prices and demands correspond to the Nash equilibrium endowments and 
in particular, Corollary \ref{cor:sharing game} and equation
\eqref{eq:Nash equilibrium price of C} imply that the
oligopolistic structure of the market leaves the prices unaffected
only when participating agents have the same risk aversion.
\begin{corollary}\label{cor:CAPM}
Assume that there exists a pair
$i,j\in\{1,...,n\}$, such that $\gamma_i\EN_i-\gamma_j\EN_j$ is not a constant. Then, the Nash and the competitive equilibrium prices of a security vector $\CC\in(\ltwo)^k$
coincide if and only if agents are homogeneous with respect to their risk aversions.
\end{corollary}
Even though the prices coincide under equal agents' risk aversions, the equilibrium allocations do not. 
We readily calculate that 
the volume in Nash equilibrium is reduced
for each security and in particular,
$\Zn_i(\ppn)=(n-1)Z_i(\pp^*)/n$, for each
$i\in\{1,...,n\}$. Recall that according to the best-demand response, each agent has motive to declare less demand for the securities she wants to buy. 
Under agents' homogeneous risk preferences, these reduced demands equilibrate at lower volumes for each security, implying that (see also Corollary
\ref{pro:quadratic equilibrium level}) a decrease in
utility surplus by a percentage equal to $(2n-1)/n^2$, for each
agent. In other words, \textit{agents suffer equal utility losses, even though the prices are equal to the competitive
ones}. 

\begin{remark}\label{rem: limits}
As in Remark \ref{rem:asymptotics}, agents with risk attitudes close to risk neutrality dominate the Nash risk-sharing transaction of any security vector $\CC$. Indeed, it holds that
$\underset{\gi\rightarrow\infty}{\lim}\Zn_i=Z_i$, for every
$\gamma_{-i}\in\mathbb{R}_+$; while
$\underset{\gamma_{-i}\rightarrow 0}{\lim}\Zn_i=Z_i$, for every
$\gi\in\mathbb{R}_+$ (where limits are taken point-wise).
\end{remark}

\subsubsection{Inefficiency and market incompleteness}\label{subsub: loss of incompletenss}

By definition, the efficiency
is higher when the tradeable securities are more correlated with
the agents' endowments. Indeed, according to Proposition \ref{pro:loss of
utilities}, under no market power, each agent suffers a
loss of utility when the market is incomplete (i.e., every
agent has a motive to complete the market). It turns out that a
similar result holds also in the case of thin risk-sharing transactions, where market power is used. 
This is established in the following
proposition (the proof of which is a matter of simple
calculations).
\begin{proposition}\label{pro:loss of Nash utilities}
In Nash price-allocation equilibrium, each individual agent suffers a
loss of utility from market incompleteness. This loss is zero
for agent $i$ if and only if $\Cn_i\in\spa(\CC)$, where $\Cn_i$ is given in
\eqref{eq: Nash contracts def}.
\end{proposition}
In other words, even when the market consists of few agents and
all use their market power, each one gets higher utility if the sharing is in a complete-market setting. 

Moreover, similarly to Proposition \ref{pro:n goes to infty},
when the number of agents goes to infinity, the differences
between the competitive and the Nash equilibrium prices and allocations vanish. This is stated in the
following proposition, where the indication $(n)$ refers to the
market of $n$ agents.
\begin{proposition}\label{pro:nC goes to infty}
Consider a sequence of agents with random endowments
$(\EN_i)_{i\in\N}$ and risk-aversion coefficients
$(\gamma_i)_{i\in\N}$. If $(\EN_i)_{i\in\N}$ is uniformly bounded in $\ltwo$ and there exist positive constants $c_l$
and $c_u$ such that $c_l\leq
\gamma_i\leq c_u$ for each $i\in\N$, then for any vector of
securities $\CC$, as $n\rightarrow \infty$ it holds that
\begin{itemize}
    \item [(i)] $||\pp^*(n)-\ppn(n)||\rightarrow 0$ and
    \item [(ii)] $||\aa_i^*(n)-\aa^{\diamond}_i(n)||\rightarrow 0$, for each
    $i\in\N$.
\end{itemize}
\end{proposition}

\subsubsection{The case of two agents}\label{subsub: CAPM two agents}
In contrast to the case of homogeneous agents, when they have different risk-aversion coefficients, the inefficiency of Nash equilibrium is not equally distributed. In fact, as in the complete-market setting, agents with sufficiently lower risk-aversion get more utility in Nash equilibrium for any security vector $\CC$. As in subsection \ref{subsec:Nash complete}, we compare utility surpluses in the simplified case of $n=2$. We calculate $\UU_1(\EN_1+\Zn_1(\ppn)\cdot(\CC-\ppn))=\UU_1(\EN_1+Z_1(\pp^*)\cdot(\CC-\pp^*))-
[(3\gamma_1-2\gamma_2)/4]\Cov(\CC,C_1^o)\cdot\Var^{-1}[\CC]\cdot\Cov(\CC,C_1^o)$, which means that agents' risk aversion is the only factor that determines
which agent benefits from the game. In particular, as in Proposition \ref{pro:more gain in Nash}, agent 1 gets more utility in Nash if and only if $\gamma_1<2\gamma_2/3$.

Regarding the price impact, comparison of \eqref{eq:PEquadr} and
\eqref{eq:Nash equilibrium price of C} gives the exact measure and
the direction of the price impact that agents' strategic
behavior causes. For each $j\in\{1,...,k\}$,
$$ p^*_j< \pn_j\,\,\,\,\,\,\text{ if and only if }\,\,\,\,\,\,\Cov(C_j,\BBn)<\Cov(C_j,\EN).$$
Hence, $\Cov(C_j,\EN-\BBn)$ can be considered as the measure of
the price impact that is caused on security $C_j$ by agents' market power (the
game creates upward price impact if $\Cov(C_j,\EN-\BBn)>0$ and
negative price impact if $\Cov(C_j,\EN-\BBn)<0$). Note that for $n=2$,
$\Cov(C_j,\EN-\BBn)=(\gamma_1-\gamma_2)\Cov(C_j,C_1^o)/2\gamma$.

This is the outcome when the predatory trading is applied by both agents in the incomplete-market setting. Recall that according to this strategy, each agent submits higher (resp.~lower) demand on the security she wants to buy (resp.~sell). This yields lower equilibrium volume, which tends to reduce agents' utility surpluses (see item (iii) of Proposition \ref{pro: best demand and endowment}). On the other hand, the game changes the equilibrium prices. In particular, assuming that agent 1 is less risk averse than agent 2, the
price impact is negative (resp.~positive) if agent 1 buys (resp.~sells)
security $C_j$.$^{31}$\footnote{$^{31}$Recall from equation
\eqref{eq:PEquadr} that the allocation of agent 1 at competitive 
equilibrium is
$\Cov(\CC,C_1^o)\cdot\Var^{-1}[\CC]$.} If $\gamma_1$ is sufficiently lower than $\gamma_2$, the beneficial price impact prevails the utility reduction that is caused by the decreased equilibrium volume. Hence, even in the incomplete markets, we may conclude that sufficiently lower risk averse agents are the dominant predatory traders. Note also, that in contrast to Brunnermeier and  Pedersen \cite{BruPed05}, where the role of a predatory trader is exogenously imposed, here which agents act as predatory traders is endogenously derived. 

\begin{remark}\label{rem:speculator Nash}
We should also point out that the higher utility surplus that sufficiently lower risk averse agents get at Nash equilibrium is independent of their risk exposure. This implies that even when the low risk averse agent is a speculator (see Remark \ref{rem:speculator}), the aforementioned game is beneficial. This is an important observation, since agents with no risk could be considered as market makers in this particular set-up. Also, it is rather common in the literature (see Liu and Wang \cite{LiWan16} and the reference therein) that market makers are assumed risk neutral (which in limiting terms implies that their risk-aversion is sufficiently low). Taking the above analysis into account, we may conclude that for lower risk averse market makers, a thin market, where all agents act strategically, could be more beneficial than a corresponding competitive one. 
\end{remark}


\bigskip

\section{The Effect of Noise Traders}\label{sec: noise trader}

The risk-sharing equilibria that we studied in the previous sections do not impose the presence of noise traders. In the related literature however, it is common (see among others Kyle \cite{Kyle89}, Palomino \cite{Pal96}) to assume that there exists some given (non-strategic) order in the market, which effectively makes the net supply different than zero. The goal of this section is to examine how the main results of the previous sections are affected by the existence of an aggregate non-zero order submitted by noise traders.

\subsection{Noise traders in the complete-market setting}\label{subsec:n complete}

We first consider the case where agents can freely design new securities to complete the risk-sharing market. In such setting, the noise traders submit a risk (random endowment) denoted by $\EN_N\in\ltwo$, which stands for the aggregate risk exposure that they want to hedge. Throughout this section we assume that $\EN_N$ is not constant and hence, the net supply in agents' risk-sharing is not zero.$^{32}$\footnote{$^{32}$An indicative example is the case of insurance companies (agents) that want to share their insurance portfolios (endowments) in a market where other firms (noise traders, clients) have submitted the risky positions ($\EN_N$) that they want to insure. Insurers satisfy the noise traders' order and at the same time share their endowments in a non-competitive manner.} 

Formally, under the additional noise traders' endowment $\EN_N$, the set of all possible sharing rules is defined as $\tilde{\A}:=\{\KK=(K_1,...,K_n)\in(\ltwo)^n:\sum_{i=1}^nK_i=\EN_N\}$ and Definition \ref{def: Patero optimal} is adapted accordingly.$^{33}$\footnote{$^{33}$Throughout this section notations with ``tilde'' refer to the features where noise traders are involved.} The generalization of the Proposition \ref{pro:equilibrium with quadratic} given below.  

\begin{proposition}\label{pro:n Pareto}
Under the presence of noise traders' aggregate endowment $\EN_N$, the unique Pareto-optimal risk-sharing is $\tilde{\KK}^o=(\tilde{K}^o_1,...,\tilde{K}^o_n)\in\tilde{\A}$, where $\tilde{K}^o_i=\nC_i^o-\npi^o_i$, for each $i\in\{1,...,n\}$, and 
\begin{equation}\label{eq:n optimal securities}
\nC_i^o=C_i^o+\lambda_i\EN_N\quad\text{and}\quad\npi^o_i=\EE[\nC_i^o]-2\gamma\Cov(\nC_i^o,\EN+\EN_N). 
\end{equation} 
\end{proposition}

In words, when agents do not behave strategically, each one gets the optimal risk-sharing security and a proportion of the noise traders' aggregate endowment; while the aggregate endowment is now $\EN+\EN_N$. 

\begin{remark}\label{rem:nt utility gain at Pareto}
Under no market power, the presence of noise traders always increases the agents' aggregate utility. Indeed, the utility of agent $i$ at the optimal transaction is $\tilde{v}_i^o=\gamma_i\Var[\nC_i^o]+\UU_i(\EN_i)$ (similarly to \eqref{eq:level at optimal sharing}) and hence $\sum_{i=1}^n\tilde{v}_i^o=\sum_{i=1}^nv_i^o+\gamma\Var[\EN_N]$. However, this does not imply that each agent's utility increases; $\tilde{v}_i^o-v_i^o=\gamma_i\Var[\lambda_i\EN_N]+2\gamma\Cov(C_i^o,\EN_N)$, which means that agent $i$ profits from the presence of noise traders when the correlation of $\EN_N$ and the payoff $C_i^o$ is non-negative. Intuitively, this is because positive  
$\Cov(C_i^o,\EN_N)$ means that agent $i$'s hedging needs are partially covered by buying some of the noise traders' endowment.    
\end{remark}

\subsubsection{Best-endowment response and Nash equilibrium}

The best-response strategy introduced in the subsection \ref{subsec:best responses} is readily generalized under the presence of noise traders. In particular, agents have agreed to apply the sharing rules \eqref{eq:n optimal securities}, for each vector of endowments that is submitted for sharing. 
	More precisely and similarly to \eqref{eq:optimal contract when B is reported}-\eqref{eq:best response problem}, if agent $i$ submits endowment $B\in\ltwo$, the security payoff she gets and the price she pays are 
\begin{equation}\label{eq:n optimal contract and price when B is reported}
    \nC_i^o(B):=\lambda_i(\EN_{-i}+\EN_N)-\lambda_{-i} B,\quad  
    \npi_1^o(B):=\EE[\nC_i^o(B)]-2\gamma\Cov(\nC_i^o(B),B+\EN_{-i}+\EN_N) .
\end{equation}
Hence, best-endowment response problem is given as $\nBr_i:={\argmax}_{B\in\ltwo}\{\nG_i(B;\EN_{-i})\}$, where $\nG_i(B;\EN_{-i}):=\UU_i(\EN_i+\nC_i^o(B)-\npi_i^o(B))$. The corresponding generalization of Proposition \ref{pro:best response} follows.
\begin{proposition}\label{pro:n best response}
For each $i\in\{1,...,n\}$, the unique (up to constants) best-endowment response of agent $i$, when the rest of the agents and noise traders have
reported aggregate endowments $\EN_{-i}$ and $\EN_N$ respectively is given by
\begin{equation}\label{eq:n best response}
  \nBr_i=\frac{1}{1+\lambda_i}\EN_i+\frac{\lambda_i^2}{1-\lambda_i^2}\left(\EN_{-i}+\EN_N\right).
\end{equation}
This gives that $\nCr_i(\nBr_i)=\nC_i^o/(1+\lambda_i)$.
\end{proposition}   

We therefore verify that the presence of an additional endowment makes agents declare exposure to $\EN_N$ at the same fashion as their motive to declare exposure to $\EN_{-i}$; while they still share only a fraction of their true endowment. The more interesting generalization, though, is regarding the Nash equilibrium endowments, (Definition \ref{def: Nash equilibrium} is adapted accordingly).
We recall from Theorem \ref{thm:nash} that $\alpha=1-\sum_{j=1}^n\lambda_j^2$.
\begin{theorem}\label{thm:n nash}
Under the presence of noise traders' endowment $\EN_N$, there exists a unique (up to constants) Nash risk-sharing
equilibrium characterized as follows: For each $i\in\{1,...,n\}$, the Nash equilibrium endowment $\nBn_i$, of agent $i$ is
\begin{equation}\label{eq:n B*i}
    \nBn_i=\lambda_{-i}\EN_i+\lambda_i^2(\nBBn+\EN_N)=\Bn_i+\frac{\lambda_i^2}{\alpha}\EN_N,
\end{equation}
where $\tilde{\BBn}:=\sum_{i=1}^n\nBn_i$ and 
\begin{equation}\label{eq:nB*}
    \nBBn=\frac{1}{\alpha}\sum_{j=1}^n\left(\lambda_{-j}\EN_j+\lambda_j^2\EN_N\right)=\BBn+\frac{1-\alpha}{\alpha}\EN_N,
\end{equation}    
and $\Bn_i$ and $\BBn$ are given in \eqref{eq:B*i} and \eqref{eq:B*}. The payoff of the Nash risk-sharing security that agent $i$ gets is
     \begin{equation}\label{eq:nC*}
    \nCn_i=\lambda_{i}(\nBBn+\EN_N)-\nBn_i=\Cn_i+\frac{\lambda_i(1-\lambda_{i})}{\alpha}\EN_N,
    \end{equation}
where $\Cn_i$ is given in \eqref{eq:C*} and the utility of agent $i$ at the Nash equilibrium is
\begin{equation}\label{eq:n level at Nash} 
v^o_i(\npin_i)=\UU_i(\EN_i+\nCn_i-\npin_i)=\gamma_i\frac{1+\lambda_i}{1-\lambda_i}\Var[\nCn_i]+\UU_i(\EN_i)=\gamma_i\frac{\gamma_i+\gamma}{\gamma_i-\gamma}\Var[\nCn_i]+\UU_i(\EN_i).
\end{equation}
\end{theorem}

As in the best-endowment response, in equilibrium each agent reports some exposure to the noise traders' endowment. In aggregate level, even in the case of equal risk aversions, agents always report aggregate exposure to $\EN_N$. Regarding the equilibrium securities, decomposition \eqref{eq:nC*} yields that each agent gets a part of $\EN_N$ and the security that she would have got without the presence of noise traders. Note also that sufficiently lower risk averse agents buy less $\EN_N$ in Nash equilibrium than in optimal risk-sharing security \eqref{eq:n optimal securities}. In other words, when $\lambda_i$ is large enough, it holds that $\lambda_i\lambda_{-i}/\sum_{j=1}^n\lambda_j\lambda_{-j}<\lambda_i$ (an inequality equivalent to inequality \eqref{eq:comparison weights}). 

\begin{remark}\label{rem:n homo}
Another consequence of the noise traders' presence is that the aggregate submitted endowment is not equal to the true aggregate endowment, even in the case of homogeneous agents. Indeed, from decompositions \eqref{eq:nB*} and \eqref{eq:nC*} and Corollary \ref{cor:sharing game}, we get that $\nBBn=\EN+\EN_N/(n-1)$ and $\nCn_i=nC^o_i/(n-1)+\EN_N/n$. The latter also implies that the volume is reduced, as in the Nash risk-sharing equilibrium with no noise traders.  
\end{remark}

\begin{remark}\label{rem:n asymptotics}
As in the case of no noise traders (see Remark \ref{rem:asymptotics}), agents whose risk preferences approach risk-neutrality dominate the game, in the sense that $\lzer$-$\underset{\gi\rightarrow\infty}{\lim}\nBn_i=\EN_i$, for every
$\gamma_{-i}\in\mathbb{R}_+$; and
$\lzer$-$\underset{\gamma_{-i}\rightarrow 0}{\lim}\nBn_i=\EN_i$, for every
$\gi\in\mathbb{R}_+$. 
\end{remark}

\subsubsection{Utility surpluses and noise traders}
 
Recall from Remark \ref{rem:nt utility gain at Pareto} that in optimal risk-sharing, agents aggregately benefit from the noise traders' endowment. Relations \eqref{eq:utility gain nash}, \eqref{eq:nC*} and \eqref{eq:n level at Nash} allow us to measure the effect of the noise traders' endowment on the utility surpluses at the Nash equilibrium. More precisely, we first calculate the difference between the aggregate utility at Nash risk-sharing equilibrium with and without the noise traders' submitted endowment, i.e.,
\begin{align}\label{eq:n gain difference}
\nonumber \sum_{i=1}^nv_i^o(\npin_i)-\sum_{i=1}^nv_i^o(\pin_i) &= \frac{1}{\alpha^2}\sum_{i=1}^n\gamma_i(1+\lambda_i)\lambda_i^2\lambda_{-i}\Var[\EN_N]+
          \frac{2}{\alpha}\sum_{i=1}^n\gamma_i(1+\lambda_i)\lambda_i\Cov(\Cn_i,\EN_N)    \\
 &= \frac{\gamma}{\alpha^2}\left(1-\sum_{i=1}^n\lambda_i^3\right)\Var[\EN_N]+\frac{2\gamma}{\alpha}\sum_{i=1}^n\lambda_i\Cov(\Cn_i,\EN_N).
\end{align}
In contrast to the optimal risk-sharing, the presence of noise traders is not always beneficial for the agents' aggregate utility. The important factor, once again, is the agents' risk-aversion coefficient. In fact, \eqref{eq:n gain difference} is always positive when $\gi=\gamma_j$, for all $i,j\in\{1,...n\}$. However, this does not necessarily holds when agents are heterogeneous with respect to their risk aversions. In particular, when the correlation between the optimal securities of relatively low risk averse agents and endowment $\EN_N$ is sufficiently lower than zero, then \eqref{eq:n gain difference} could be negative. Intuitively, when payoff $\EN_N$ is of opposite direction than the hedging needs of lower risk averse agents, the aggregate utility is reduced. 

As before, we examine the individual utility surplus at Nash equilibrium in the special case of $n=2$. We first state the following consequence of Theorem \ref{thm:n nash}.
\begin{corollary}\label{cor:n gains}
Let $n=2$ and assume the presence of noise traders' endowment $\EN_N$. Then
\[v_1^o(\npi^o_1)-v_1^o(\npin_1)= \frac{3\gamma_1-2\gamma_2}{4}\Var[C_1^o]+\Var[\EN_N]\left(\frac{\gamma^2}{\gamma_1}-\frac{\gamma_1+2\gamma_2}{4}\right)
+2\Cov(C_1^o,\EN_N)\left(\gamma-\frac{\gamma_1+2\gamma_2}{4}\right),\]
where $C_1^o$ is given in \eqref{eq:C*}. In particular,  $\underset{\gamma_1\rightarrow 0}{\lim}\left[v_1^o(\npi^o_1)-v_1^o(\npin_1)\right]=-\gamma_2\Var[\EN_2+\EN_N]/2$. 
\end{corollary}

Therefore, as long as $\EN_2-\EN_N$ is not a constant, the presence of the noise traders does not violate the general result that \textit{Nash equilibrium is beneficial for sufficiently lower risk averse agents, who eventually act as predatory traders}. However, in contrast to Proposition \ref{pro:more gain in Nash}, the condition $\gamma_1<2\gamma_2/3$ is not sufficient to yield more utility surplus for agent 1 in Nash equilibrium. Indeed, as covariance $\Cov(C_1^o,\EN_N)$ gets lower (which is unpleasant for agent 1), the risk-aversion level under which the Nash equilibrium is more profitable for agent 1 decreases too. In other words, the noise traders reduce the profits from the predatory trading. Recall that predatory traders in our model are still risk averse with specific hedging needs. If for example $\Cov(C_1^o,\EN_N)$ is negative, the noise traders' endowment deteriorates the partial hedging of agent 1 that is achieved through the risk-sharing. Corollary \ref{cor:n gains} verifies, however, that Nash equilibrium is always beneficial for sufficiently low $\gamma_1$, no matter the sign and the size of $\Cov(C_1^o,\EN_N)$.  

\begin{remark}\label{rem:n gains of noise traders}
The difference in agents' risk aversion is also an important factor for the aggregate cash compensation that the noise traders have to pay. In particular, when $\gamma_i\neq\gamma_j$ for some $i,j\in\{1,...,n\}$, it is possible that noise traders (together with the sufficiently risk tolerant agents) benefit from the Nash equilibrium$^{34}$\footnote{$^{34}$This result is in the line with the corresponding results in Palomino \cite{Pal96}.}. For this, we calculate the difference of cash amounts paid by the noise traders to hedge position $\EN_N$ in two equilibria, i.e.,
\[\npi^o_N-\npin_N=2\gamma\Cov(\EN_N,\BBn-\EN)+2\gamma\frac{\sum_{i=1}^n\lambda_i^2}{1-\sum_{i=1}^n\lambda_i^2}\Var[\EN_N],\]
which implies that when $\Cov(\EN_N,\BBn-\EN)$ is sufficiently lower than zero, the price impact caused by the Nash game works in noise traders' favor. Note that for homogeneous agents, we have that $\BBn=\EN$, meaning that noise traders pay always more in Nash equilibrium.
\end{remark}

\subsection{Noise traders in the incomplete-market setting}\label{subsec:n incomplete}

With the vector of tradeable securities being fixed, we model the presence of noise traders following the related literature (see e.g.~Kyle \cite{Kyle89}). More precisely, given a security vector $\CC\in(\ltwo)^k$, noise traders submit an aggregate order (a vector $\zeta\in\R^k \backslash \{0\}$), that could be satisfied at any price level. In other words, the market clearing condition becomes $\sum_{i=1}^nZ_i(\pp)+\zeta=0$, where $Z_i$ is the demand function of agent $i$. Following the arguments of Proposition \ref{pro:equilibrium with quadratic}, we get that the competitive equilibrium price of $\CC$ is 
\begin{equation}\label{eq:n PEquadr}
    \npp^*=\EE[\CC]-2\gamma\Cov(\CC,\EN)+2\gamma\zeta\cdot\Var[\CC]=\pp^*+2\gamma\zeta\cdot\Var[\CC]
\end{equation}
and the (unique) equilibrium allocation for agent $i$ is
\begin{equation}\label{eq:n DEquadr}
    \naa_i^*=\Cov\left(\CC,C_i^o\right)\cdot\Var^{-1}[\CC]+\lambda_i\zeta=\aa_i^*+\lambda_i\zeta,
\end{equation}
for each $i\in\{1,...,n\}$, where $\pp^*$ and $\aa_i^*$ are given in \eqref{eq:PEquadr}. In words, as in complete-market setting, each agent gets the proportion $\lambda_i$ of the submitted order of the noise traders and the equilibrium prices change accordingly (increase when noise traders want to buy and decrease otherwise). 
\begin{remark}\label{eq:n gains in Pareto}
A direct consequence of the fact that the noise traders' order could be satisfied at any price is that agents' aggregate utility is increased. Indeed, $\sum_{i=1}^n\UU_i(\EN_i+\naa_i^*\cdot\CC-\naa_i^*\cdot\npp^*)$= $\sum_{i=1}^nv_i(\pp^*;\CC)+\gamma\zeta\cdot\Var[\CC]\cdot\zeta$. Similarly to the discussion in Remark \ref{rem:nt utility gain at Pareto}, however, it is not necessarily true that each agent enjoys higher utility when the additional order $\zeta\neq 0$ is considered. Simple calculations yield that the effect of the noise traders on agent $i$'s utility is positive if and only if 
$\lambda_i\zeta\cdot\Var[\EN_N]\cdot\zeta>2\zeta\cdot\Cov(C_i^o,\CC)$. Intuitively, this occurs when the noise traders' order has low correlation with the security agent $i$ wants to buy, meaning that she hedges some of her risk exposure by satisfying a part of noise traders' order (recall also the analogous comment in Remark \ref{rem:nt utility gain at Pareto}).
\end{remark}

\subsubsection{Best-demand response and Nash equilibrium}

The best-demand response of an individual agent follows the line of subsection \ref{subsec:best response constrained}. The only difference is that the aggregate demand that agent $i$ faces is now $\sum_{j\neq
i}Z_j(\pp)+\zeta=\left[(\EE[\CC]-\pp)/2\gamma_{-i}-\Cov(\CC,\EN_{-i}-\zeta\cdot\CC)\right]\cdot\Var^{-1}[\CC]$. Therefore, the maximization problem \eqref{eq:utility given Zi} is adapted accordingly, i.e., the best price of agent $i$ is the maximizer of $\phi_i(\pp;Z_{-i}(\pp)+\zeta)$.  
We readily get the following results. 
\begin{proposition}\label{pro:n best response ini} 
The equilibrium price of a vector of securities $\CC\in(\ltwo)^k$ that maximizes the utility of agent
$i$, given the other agents' aggregate demand function and noise traders' aggregate order $\zeta\in\R^k\backslash\{0\}$, is 
\begin{equation}\label{eq:n best price}
    \nppr_i= \ppr_i+\frac{2\gamma}{1-\lambda_i^2}\zeta\cdot\Var[\CC].
\end{equation}
where $\ppr_i$ is given in \eqref{eq:best price}. Furthermore, agent $i$'s best-demand response is  $\nZr_i(\pp)= \Zr_i(\pp)+\zeta\lambda_i^2/(1-\lambda_i^2)$, where $\Zr_i(\pp)$ is given in Proposition \ref{pro: best demand and endowment}. In particular, $\nZr_i(\nppr_i)=Z_i(\npp^*)/(1+\lambda_i)$.
\end{proposition}

The presence of noise traders shifts the demand function of an agent in the direction of their order. In fact, the demand function $\nZr_i$ coincides with the demand function that is associated with endowment $\Br_i-\lambda_i^2\zeta\cdot\CC/(1-\lambda_i^2)$, where $\Br_i$ is the best-endowment response without the presence of noise traders, given in \eqref{eq:best response}. In order words, agent $i$, acting as predatory trader, has motive to declare some exposure to the tradeable securities $\CC$ in a way that increases the aggregate demand of the securities that the noise traders want to buy. 

For the Nash price-demand equilibrium, we slight change the Definition \ref{def: Nash equilibrium of C} in that we call a pair $(\nppn,\tilde{\mathbf{Z}}^{\diamond})\in\R^k\times\ZZ$, Nash price-demand equilibrium under an addition noise traders' order $\zeta\neq 0$, if $\sum_{i=1}^n\nZn_i(\nppn)+\zeta=0$, and that for each $i\in\{1,...,n\}$, $\phi_i(\nppn;\nZn_{-i}(\nppn)+\zeta)\geq\phi_i(\pp;\nZn_{-i}(\pp)+\zeta)$, for all $\pp\in\R^k$.

\begin{theorem}\label{thm:n Nash equilibrium price}
For every vector of securities $\CC\in(\ltwo)^k$, and noise traders' aggregate order $\zeta\in\R^k\backslash\{0\}$, there exists a unique Nash price-demand equilibrium $(\nppn,\tilde{\mathbf{Z}}^{\diamond})\in\R^k\times\ZZ$ given by 
\begin{equation}\label{eq:n Nash equilbrium price of C}
    \nppn=\EE[\CC]-2\gamma\Cov\left(\CC,\BBn-\frac{\zeta\cdot\CC}{\alpha}\right)=\ppn+\frac{2\gamma}{\alpha}\zeta\cdot\Var[\CC],
\end{equation}
and for each $i\in\{1,...,n\}$
\begin{equation}\label{eq:n Nash demands}
        \nZn_i(\pp)=\left[\frac{\EE[\CC]-\pp}{2\gi}-\Cov\left(\CC,\Bn_i-\frac{\lambda_i^2}{\alpha}\zeta\cdot\CC\right)\right]\cdot\Var^{-1}[\CC]=
        \Zn_i(\pp)+\frac{\lambda_i^2}{\alpha}\zeta,
\end{equation}
where $\ppn$ and $\Zn_i$ are given in \eqref{eq:Nash equilibrium price of C} and \eqref{eq:Nash demands}. Finally, the induced Nash allocation is
\begin{equation}\label{eq:n Nash equilbirium allocation}
    \nZn_i(\nppn)=\Zn_i(\ppn)-\frac{\lambda_i(1-\lambda_i)}{\alpha}\zeta.
\end{equation}
\end{theorem}
 
As in the best-demand response, the equilibrium demands are shifted in the direction of the noise traders' order, reflecting the fact that each agent declares exposure to the tradeable securities. The more risk tolerant an agent is, the more exposure on $\CC$ she submits. Regarding the Nash allocation, all agents satisfy some of the noise traders' order, however the more risk tolerant ones get a smaller part. In addition, decomposition \eqref{eq:n Nash equilbirium allocation} implies that (as in the transaction without the noise traders' order) the Nash equilibrium has less volume than the competitive one. 
\begin{remark}\label{rem:n homo incomplete}
In contrast to Corollary \ref{cor:CAPM}, equality on the agents' risk aversion does not imply equality of Nash and competitive equilibrium prices, when noise traders submits a non-zero order. The necessary and sufficient condition for $\nppn=\ppn$ is that $\Cov(\CC,\EN-\BBn)=(1-\alpha)\zeta\cdot\Var[\CC]/\alpha$, which does not holds when agents are homogeneous with respect to their risk aversions and $\zeta\neq 0$.  
\end{remark}

\subsubsection{Utility surpluses and noise traders}
 
As in previous sections, for the individual utility at Nash equilibrium and its comparison to the competitive ones, we consider the special case of $n=2$. We first state the following consequence of Theorem \ref{thm:n nash}.
\begin{corollary}\label{cor:n gains incomplete}
Let $n=2$ and assume the presence of noise traders' order $\zeta\in\R^k\backslash\{0\}$. Then,
\begin{equation*}
\UU_1(\EN_1+Z_1(\npp^*)\cdot(\CC-\npp^*))-\UU_1(\EN_1+\nZn_1(\nppn)\cdot(\CC-\nppn))=\quad\quad\quad\quad\quad\quad\quad\quad
\end{equation*}
\begin{align}\label{eq:n gain n=2 incomplete}
=& \,\frac{3\gamma_1-2\gamma_2}{4}\Cov(\CC,C_1^o) \cdot\Var^{-1}[\CC]\cdot\Cov(\CC,C_1^o) +\left(\frac{\gamma^2}{\gamma_1}-\frac{\gamma_1+2\gamma_2}{4}\right)\zeta\cdot\Var[\CC]\cdot\zeta +\\
\nonumber & +2\zeta\cdot\Cov(\CC,C_1^o) \left(\gamma -\frac{\gamma_1+2\gamma_2}{4}\right),
\end{align}
where $C_1^o$ is given in \eqref{eq:C*}. In particular, the limit of utility difference \eqref{eq:n gain n=2 incomplete} as $\gamma_1$ goes to zero equals to  
$-(\gamma_2/2)\Cov(\CC,\zeta\cdot\CC+\EN_2) \cdot\Var^{-1}[\CC]\cdot\Cov(\CC,\zeta\cdot\CC+\EN_2)$.
\end{corollary}

Note that the first term in \eqref{eq:n gain n=2 incomplete} is the utility difference at competitive and Nash equilibria without the noise traders' order $\zeta$. Hence, similarly to Corollary \ref{cor:n gains}, as long as $\EN_2+\zeta\cdot\CC$ is not a constant, the presence of the noise traders does not violate the general result that Nash equilibrium is beneficial for sufficiently lower risk averse agents. However, in contrast to case without noise traders (see subsection \ref{subsub: CAPM two agents}), the condition $\gamma_1<2\gamma_2/3$ is not sufficient to yield more utility surplus for agent 1 in Nash equilibrium. Indeed, as the term $\zeta\cdot\Cov(\CC,C_1^o)$ gets higher (which is unpleasant for agent 1), the risk aversion level under which the Nash equilibrium is more profitable for agent 1 decreases. Corollary \ref{cor:n gains incomplete} verifies, however, that Nash equilibrium is always beneficial for sufficiently low $\gamma_1$ no matter the sign and the size of $\zeta\cdot\Cov(\CC,C_1^o)$.  

\begin{remark}\label{rem:n gains of noise traders incomplete}
As expected from Remark \ref{rem:n gains of noise traders}, it holds also in incomplete-market setting that the aggregate cash paid by the noise traders is heavily influenced by agents' risk-aversion coefficients. For this, we observe that $\nppn\cdot\zeta-\npp^*\cdot\zeta=(\ppn-\pp^*)\cdot\zeta+2\gamma(1-\alpha)\zeta\cdot\Var[\CC]\cdot\zeta/\alpha$,
which implies that when the price impact caused by Nash game and order $\zeta$ have opposite directions, the total cash amount that noise traders pay in Nash equilibrium may be less than in competitive transaction. Note that when $\gamma_i=\gamma_j$, it holds that $\ppn=\pp^*$ and hence noise traders pay always more in Nash equilibrium.
\end{remark}

\bigskip

\section{Difference in Beliefs}\label{sec:difference}

So far we have assumed that agents have the same beliefs regarding the probability of future outcomes (probability measure $\PP$ is assumed common among agents). In practice however, agents do have different probability assessments for the probability distributions of the tradeable securities. 
It is therefore important to examine whether and how the main findings of Sections \ref{sec:Agent's Strategic Behavior in Risk Sharing Market} and \ref{sec:nash equilibria} are affected when model allows difference in agents' beliefs. In the incomplete-market setting, we adapt a model similar to the one of Kyle et al.~\cite{KylObiWan14} and Rostek and Weretka \cite{RosWer12}, where the agents' assessments differ in the expected value of the payoffs and their covariances.$^{35}$\footnote{$^{35}$When agents with M-V preferences trade a given vector of securities, their beliefs' differences are incorporated in the expectations and variances-covariances. When, however, the market setting is complete, for the designing of the Pareto-optimal securities (therefore for the induced Nash game) one should take into account the agents' differences on the probability measures on whole $\Omega$. A game on this general set-up, which is beyond the scope of this paper, is recently established in Anthropelos and Kardaras \cite{AnthKar15} (see also the discussion in subsection \ref{subsec:db complete} below).}
Agents still declare different risk exposure than their true endowments, and in fact they have motive to report different beliefs too. As before, sufficiently lower risk averse agents get more utility, however different beliefs do not necessarily imply that the predatory trading is more profitable. As explained below, the predatory trader's utility surplus decreases when the average agents' assessments give higher value to the securities she buys at equilibrium. 

\subsection{Different beliefs in the incomplete-market setting}\label{subsec:db incomplete}

As in  Kyle et al.~\cite{KylObiWan14} and Rostek and Weretka \cite{RosWer12}, we assume that agents have common assessments for variance-covariance matrix of $\CC$; while they possibly have different assessments for the expectation of vectors $\EEE$, $\CC$ and their covariances. The utility functional of agent $i$ takes the form: $\UU_i(X)=\EE_i[X]-\gi\Var_i[X]$ for any security payoff $X$, where $\EE_i$ and $\Var_i$ denotes the expectation and variance under the subjective probability measure of agent $i$. Hence, agent $i$'s demand function is
$Z_i(\pp)=\left[(\EE_i[\CC]-\pp)/2\gi-\Cov_i(\CC,\EN_i)\right]\cdot\Var^{-1}[\CC]$ (see also \eqref{eq:quadr demand}), where $\Cov_i$ stands for the covariance map under agent $i$'s beliefs. To facilitate the reading, we denote the intercept points of agents' demand function as \[\R^k\ni\ini_i:=\frac{\EE_i[\CC]}{2\gi}-\Cov_i(\CC,\EN_i).\] 
It is important to note that $\ini_i$ reflects not only the hedging needs of agent $i$, but also her beliefs on the tradeable securities and their covariances with her endowment. We also define $\ini:=\sum_{i=1}^n\ini_i$ and $\ini_{-i}:=\sum_{j\neq i}\ini_j$.

Under different beliefs and with the aforementioned notation, the generalizations of Proposition \ref{pro:equilibrium with quadratic} and Corollary \ref{pro:quadratic equilibrium level} give that the (unique) competitive price-allocation equilibrium of $\CC$ becomes
\begin{equation}\label{eq:db optimal price}
\dpp=2\gamma\ini\quad\text{and}\quad\daa_i=(\ini_i-\lambda_i\ini)\cdot\Var^{-1}[\CC],^{36}\footnote{$^{36}$The notations with ``hat" refer to features of the model that allows different beliefs.}
\end{equation}
for each $i\in\{1,...,n\}$, while her utility at equilibrium is
\begin{equation}\label{eq:db gain at optimal}
v_i(\dpp;\CC)=\gi Z_i(\dpp)\cdot\Var[\CC]\cdot Z_i(\dpp)+\UU_i(\EN_i)= \gi(\ini_i-\lambda_i\ini)\cdot\Var^{-1}[\CC]\cdot(\ini_i-\lambda_i\ini)+\UU_i(\EN_i).
\end{equation}
\begin{remark}\label{rem:db volume}
It follows from \eqref{eq:quadratic equilibrium level} and \eqref{eq:db gain at optimal}, that the more an agent trades at equilibrium the more utility surplus she gets. While in aggregate level different beliefs implies higher utility$^{37}$\footnote{$^{37}$The fact that difference of beliefs gives rise to more beneficial trading opportunities among agents has also been highlighted in the literature (see among others Kyle et al.~\cite{KylObiWan14} and the references therein). Intuitively, the agents' agreement to disagree on their beliefs increases the volume of the equilibrium transaction and hence their gain from trading. In a sense, through the trading of $\CC$, agents (partially) share not only their risks, but also their beliefs' heterogeneity.}, for each individual agent different beliefs does not always imply higher volume and hence higher utility surplus. Let for example $k=1$ and further assume that 
$\Cov_i(C,\EN_j)=\Cov_j(C,\EN_j)$, for all $j\neq i$. Simple calculations yield that agent $i$'s utility surplus at competitive equilibrium is lower under common beliefs if and only if
\begin{equation}\label{db gains comparison 3}
\left(\sum_{j\neq i}\frac{\EE_j[C]}{\lambda^{-i}_j}-\EE_i[C]\right)\Cov_i(C,C_i^o)<
\frac{1}{4(\gi+\gamma_{-i})}\left(\sum_{j\neq i}\frac{\EE_j[C]}{\lambda^{-i}_j}-\EE_i[C]\right)^2,
\end{equation}
where $\lambda^{-i}_j:=\gamma_{-i}/\gamma_{j}$. Recall from \eqref{eq:PEquadr} that $\Cov_i(C,C_i^o)>0$ means that agent $i$ buys $C$ at equilibrium. Therefore, difference in beliefs is beneficial for agent $i$, if the scaled aggregate expectations of $C$ submitted by the rest of agents is lower (resp.~higher) than her expectation, when agent $i$ buys (resp.~sells) security $C$. However, \eqref{db gains comparison 3} may also hold when $\Cov_i(C,C_i^o)$ is sufficiently large even if $\sum_{j\neq i}\EE_j[C]/\lambda^{-i}_j>\EE_i[C]$.
\end{remark}


\subsection{Best-demand response and Nash equilibrium under different beliefs}\label{subsec:db best response}
We now adapt the arguments of subsection \ref{subsec:best response constrained} and take the position of agent $i$, who knows the demand of the rest of the agents. Under agents' different beliefs, agent $i$, acting as predatory trader, may exploit not only the hedging needs submitted by the other agents, but also their aggregate reported beliefs. More precisely, given the aggregate demand $Z_{-i}(\pp)=\left(\ini_{-i}-\pp/2\gamma_{-i}\right)\cdot\Var^{-1}[C]$, (recall that $\gamma_{-i}=(\sum_{j\neq i}1/\gamma_j)^{-1}$) the best equilibrium price $\dppr_i$ for agent $i$ is the solution of the corresponding version of optimization problem \eqref{eq:utility given Zi}.

\begin{proposition}\label{pro:df best-response price and demand}
Under possibly different beliefs, the equilibrium price of a vector of securities $\CC\in (\ltwo)^k$ that maximizes the utility of agent $i$, given the aggregate demand function of the other agents, is
\begin{equation}\label{eq:d best price}
\dppr_i=\frac{2\gamma}{1+\lambda_i}\left(\ini_i+\frac{\ini_{-i}}{\lambda_{-i}}\right).
\end{equation}
Her best-demand response is $\dZr_i(\pp)= \left[(\inir_i-\pp)/2\gi\right]\cdot\Var^{-1}[\CC]$,
where $\inir_i:=\ini_i/(1+\lambda_i)+\lambda^2_i\ini_{-i}/(1-\lambda^2_i)$, and the implementation of the best-demand response results in the following utility increase 
\begin{equation}\label{eq:d best demand utility gain}
\UU_i  \left(\EN_i+\dZr_i(\dppr_i)\cdot(\CC-\dppr_i)\right) -\UU_i\left(\EN_i+Z_i(\dpp)\cdot(\CC-\dpp)\right)=\frac{\lambda^2_{i}}{1-\lambda_i^2}v_i(\dpp;\CC).
\end{equation} 
\end{proposition}

Therefore, each agent has motive to report demand that departs not only from her true hedging needs, but also from her true beliefs. In particular, we may write the best-demand response as
\begin{equation}\label{eq:db demand}
\dZr_i(\pp)=\left(\frac{\hat{\EE}_i[\CC]-\pp}{2\gi}-\widehat{\Cov}_i(\CC,\Br_i)\right)\cdot\Var^{-1}[\CC],
\end{equation}
where $\hat{\EE}_i[\CC]$ can be considered as the reported subjective expectation of agent $i$ and $\widehat{\Cov}_i(\CC,\Br_i)$ her reported covariance vector ($\Br_i$ is the best-endowment response given in \eqref{eq:best response}). Simple calculations yield that
\begin{align}
\hat{\EE}_i[\CC]&=\frac{\EE_i[\CC]}{1+\lambda_i}+\frac{\lambda_i}{1+\lambda_i}\sum_{j\neq i}\frac{\EE_j[\CC]}{\lambda_j^{-i}}\quad\text{and}\label{eq:db demand expect}\\
\widehat{\Cov}_i(\CC,\Br_i)& = \Cov_i\left(\CC,\Br_i\right)+
\frac{\lambda^2_i}{1-\lambda^2_i}\left(\sum_{j\neq i}\Cov_j(\CC,\EN_j)-\Cov_i\left(\CC,\EN_{-i}\right)\right).\label{eq:db demand covariance}
\end{align}

Representation \eqref{eq:db demand} gives the decomposition of the best-demand response to its endowment and beliefs parts. 
It follows from \eqref{eq:db demand expect} that agent $i$ reports an average of her own beliefs and the \textit{weighted average beliefs} of the rest of the agents (scaled by their relative risk tolerance).
On the other hand, representation \eqref{eq:db demand covariance} gives the shift of the reported covariance that is caused by the difference in beliefs only on covariances. 

We clearly get from \eqref{eq:d best demand utility gain} that the gains from this predatory strategy increase when agent is more risk tolerant and when the utility surplus at the competitive allocation is higher (exactly as in the case of common beliefs stated in the third item of Proposition \ref{pro: best demand and endowment}). Taking also Remark \ref{rem:db volume} into account, it follows that it is not necessarily true that the predatory trading is more profitable when agents have different beliefs. The reason for this is apparent in \eqref{db gains comparison 3}; when average beliefs of the other agent gives higher (resp.~lower) expectations on the securities that the predatory agent wants to buy (resp.~sell), then the benefits from the predatory trading are reduced.



Nash price-demand equilibrium under different beliefs follows the same line as in Theorem \ref{thm:Nash equilibrium price}. 
\begin{theorem}\label{thm:db Nash equilibrium price}
Under possibly different beliefs, for every vector of securities $\CC\in(\ltwo)^k$, there is a
unique Nash price-demand equilibrium $(\dppn,\mathbf{\dZn})\in\R^k\times\ZZ$ given by
\begin{equation}\label{eq:db Nash demands}
  \dppn=2\gamma\sum_{i=1}^n\inin_i\quad\text{and}\quad        \dZn_i(\pp)=\left(\inin_i-\frac{\pp}{2\gi}\right)\cdot\Var^{-1}[\CC],\quad\text{for each }i\in\{1,...,n\},
\end{equation}
where 
\begin{equation}\label{eq:db Nash demands ini}
        \inin_i=\lambda_{-i}\ini_i+\frac{\lambda_i^2}{\alpha}\sum_{j=1}^n\lambda_{-j}\ini_j.
\end{equation}
\end{theorem}

The departure of the Nash equilibrium demands from the true ones stems from the submission of different hedging needs and different beliefs. Comparison of \eqref{eq:db Nash demands ini} with \eqref{eq:B*i} and \eqref{eq:B*} implies that the submitted demands follow the same structure of the Nash endowments. In particular, the coefficient in front of $\ini_i$ in \eqref{eq:db Nash demands ini} is $\lambda_{-i}+\lambda_i^2\lambda_{-i}/\alpha$, which is strictly less than 1 for each $i$, meaning that agent $i$ under-reports both her true risk exposure and beliefs (as the latter are incorporated in parameter $\ini_i$). This behavior gets clearer if we assume that agents agree on covariance assessments. Then, 
$$\inin_i=\frac{\lambda_{-i}(1+\lambda_i^2)}{\alpha}\EE_i[\CC]+\frac{\lambda_i^2}{\alpha}\sum_{j\neq i}\lambda_{-j}\EE_j[\CC] -\Cov(\CC,\Bn_i),$$
which gives a clear decomposition of the reported beliefs and hedging needs. 

\subsubsection{The case of homogeneous agents}\label{subsub:db homo}

As we have pointed out before, the special case where $\gamma_i=\gamma_j$, for all $i,j\in\{1,...,n\}$ is widely common in the literature and, as we have seen in Corollary \ref{cor:CAPM}, it is the only case where the game leaves the equilibrium prices unaffected. Even in this case however the equilibrium volume is reduced. Below, we examine how agents' different beliefs affect these results. 
\begin{corollary}\label{cor:db homo}
Let agents have the same risk aversions but different beliefs. Then, the Nash and the competitive equilibrium price of a vector of securities $\CC\in(\ltwo)^k$ coincide, the equilibrium volume is reduced and in fact $\dZn_i(\dppn)=(n-1)Z_i(\dpp)/n$, for each
$i\in\{1,...,n\}$. 
\end{corollary}

\begin{remark}\label{rem:db homo}
The only difference with Corollary \ref{cor:CAPM} is that equality of prices $\dpp=\dppn$ does not necessarily imply that agents have the same risk aversion.  
\end{remark}

\subsubsection{Utility surpluses and different beliefs}\label{subsub:db two agents}
As in previous sections, for the individual utility at Nash equilibrium and its comparison to the competitive one, we consider the special case of $n=2$. We first state the following consequence of Theorem \ref{thm:db Nash equilibrium price}.
\begin{corollary}\label{cor:db more gain in Nash}
Let $n=2$ and consider agents with possibly different beliefs. The utility of agent 1 at the Nash price-demand equilibrium of a vector of securities $\CC\in(\ltwo)^k$ is 
\begin{equation}\label{eq:db gain at Nash}
v_1(\dppn;\CC)=\gamma_1 \dZn_1(\dppn)\cdot\Var[\CC]\cdot \dZn_1(\dppn)+\UU_1(\EN_1)= \frac{1+\lambda_1}{4(1-\lambda_1)} v_1(\dpp;\CC).
\end{equation}
In particular, agent 1 gets more utility surplus at Nash price-demand equilibrium than in the competitive equilibrium if and only if $\gamma_1<2\gamma_{2}/3$. 
\end{corollary}

In other words, as in subsection \ref{subsub: CAPM two agents} and
Proposition \ref{pro:more gain in Nash}, agents with sufficiently higher risk tolerance prefer the Nash game, since the resulting utility is always higher, for any vector of tradeable securities. Furthermore, the utility surplus that she enjoys in Nash equilibrium increases with the volume of the competitive allocation. As discussed in Remark \ref{rem:db volume}, difference in beliefs does not necessarily imply higher equilibrium allocation for each agent, which consequently means that difference in beliefs does not always imply higher utility surplus for the predatory trader (i.e., difference in beliefs does not necessarily create more opportunities for predatory trading).

\subsection{The special case of complete-market setting}\label{subsec:db complete}

Under different beliefs, the complete-market setting is more convoluted. In general, different beliefs mean that agents have different subjective probability measures on whole $(\Omega,\FF)$. Therefore, the completion of the market needs the design of securities that not only share the endowments, but also share the deviations of agents' probability measures. This implies, in particular, that the linear sharing rules (as the ones of Proposition \ref{pro:optimal sharing}) could not give optimal sharing. This problem has recently been studied in Anthropelos and Kardaras \cite{AnthKar15} (see Theorem 1.2), where exponential utility is imposed and thus agents' utility functional coincides with M-V preferences under normal distributed payoffs. Although, the study of Nash equilibria under general difference in beliefs is beyond the scope of this paper, we are able to see how the complete-market setting is affected when agents agree to disagree only on the expectations of their endowments. This is a just a special case of Theorem \ref{thm:db Nash equilibrium price} when $\text{span}(\CC)=\text{span}(\EEE)$.

\bigskip

\bigskip

\section{Conclusion}\label{sec:conclusion}

The article establishes a novel way to model agents' strategic behavior in thin risk-sharing transactions,
under symmetric bargaining power and information structure. In contrast to the relevant literature, the way agents exploit their market power stems from their motive to declare risk exposures different than their true random endowments. Our analysis includes two market settings, the
complete where agents design new securities, and the incomplete
where they negotiate the transaction of a given vector of
tradeable securities. In both settings, the risk-sharing
equilibrium is an outcome of a pure-strategy Nash
game.

In the complete case, agents negotiate the designing of the risk-sharing by proposing securities and pricing kernels that are consistent with the
optimal sharing rules. Since the sharing rules are functions of
the submitted endowments, reporting specific random endowments in the sharing
mechanism is equivalent to proposing securities and pricing kernel. It is shown that each agent has motive to report less
risk exposure to her true endowment and motive to declare an additional position proportional
to the aggregate endowment reported by the other agents. Under M-V preferences, the Nash equilibrium
does exist, is unique and in any non-trivial case is different
than the optimal risk-sharing transaction. Although the Nash game
implies a loss in the aggregate utility, for the sufficiently lower
risk averse agents the surplus of utility is higher at Nash
equilibrium than in the Pareto optimal one. Therefore, when the
market is thin, agents with sufficiently lower risk aversion could be considered as predatory traders who not only reduce the 
aggregate utility, but also absorb utility surpluses from their counterparties.

The situation is similar in the incomplete-market setting, where
agents negotiate the price and allocation of a given vector of
securities. Each agent optimally responds to the other agents'
orders by submitting the appropriate demand function that clears
out the market at the price that maximizes her own utility. The fixed
point of this negotiation is defined as the Nash price-allocation equilibrium. In principle, this market structure can
be considered as an oligopolistic version of the CAPM. Again, the existence and uniqueness of the Nash equilibrium is proved to hold and a clear connection between the complete and incomplete market setting is established. Namely, the submitted demand functions in Nash
equilibrium reflect the same hedging needs as the agents'
Nash equilibrium endowments. Moreover, the Nash and
the competitive equilibrium prices coincide if and only if all participating
agents have the same risk aversion (even in this case however the
volume is lower and the allocation is inefficient). As in the
complete market, when agents are not homogeneous, the price impact caused by the Nash game is beneficial for relatively low risk averse agents, who eventually play the role of predatory traders. 

The main results of the paper also hold when the model allows noise traders (i.e., non-zero net supply) or heterogeneous agents' beliefs (the latter, examined in the incomplete-market setting). In particular, under an additional order by noise traders or under different agents' beliefs, sufficiently lower risk averse agents still get more utility at Nash. However, the benefits of the predatory trading are not necessarily increased in these generalized models.

\bigskip

\bigskip

\appendix

\section{Proofs}\label{sec:proofs}
\subsection*{Proof of Proposition \ref{pro:optimal sharing}}
Thanks to cash-invariance property of the M-V preferences and the
assumed zero supply, the optimal-sharing securities are the ones
that maximize the sum of agents' utility (see among others
Acciaio \cite{Acc07}, Jouini et al.~\cite{JouSchTou08}). We have that for every
$\CC=(C_1,...,C_n)\in\A$
$$\sum\limits_{i=1}^n\UU_i(\EN_i+C_i)=\sum\limits_{i=1}^n\EE[\EN_i]-\sum\limits_{i=1}^n\gi\Var[C_i+\EN_i].$$
Hence, it is enough to find $\CC\in\left(\ltwo\right)^n$
that minimizes the sum $\sum\limits_{i=1}^n\gi\Var[C_i+\EN_i]$.
Note the for each $i\in\{1,...,n\}$,
$$C_i^o+\EN_i=\frac{\gamma}{\gi}\EN,$$
where $C_i^o$ is defined as the payoff $\aa_i^o\cdot\EEE$,  and
$\gi\Var[C^o_i+\EN_i]=\frac{\gamma^2}{\gi}\Var[\EN],$
where we recall that $\EN=\sum_{i=1}^n\EN_i$. Therefore,
$$\sum\limits_{i=1}^n\gi\Var[C^o_i+\EN_i]=\sum\limits_{i=1}^n\frac{\gamma^2}{\gi}\Var[\EN_i]=\gamma\Var[\EN].$$
It is then enough to show that
$\sum\limits_{i=1}^n\gi\Var[C_i+\EN_i]\geq\gamma\Var[\EN],$ for
all $\CC\in\A$, or equivalently
\begin{equation}\label{eq:optimal quadratic rule enough 2}
\sum\limits_{i=1}^n\gamma\gi\Var[C_i+\EN_i]\geq\gamma^2\Var[\EN],\quad\text{for
all }\CC\in\A.
\end{equation}
The left hand side of \eqref{eq:optimal quadratic rule enough 2}
equals to
$$\sum\limits_{i=1}^n\frac{\gamma}{\gi}\gi^2\Var[C_i+\EN_i]=\sum\limits_{i=1}^n\frac{\gamma}{\gi}\Var[\gi(C_i+\EN_i)]$$
and inequality \eqref{eq:optimal quadratic rule enough 2} follows
by the convexity of $\Var[\cdot]$ and the fact that
$\sum_{i=1}^nC_i=0$.

The fact that the equilibrium price of endowments are given by
$\ppi^o=\EE[\EEE]-2\gamma\mathbf{1}_n\cdot\Var[\EEE]$ follows from
Proposition \ref{pro:equilibrium with quadratic}.

\subsection*{Proof of Proposition \ref{pro:best response}}
We fix vector
$(\EN_1,...,\EN_{i-1},\EN_{i+1},...,\EN_n)\in(\ltwo)^{n-1}$
and from
 \eqref{eq:utility when B is reported} (with a slight abuse of notation) we get
that
\begin{eqnarray}\label{eq:utility when B is reported proof}
  \nonumber G_i(B) &=& \EE[\EN_i]-\frac{\gamma^2}{\gi}\Var[\EN]-\frac{(\gi-\gamma)^2}{\gi}\Var[\EN_i-B]-
\frac{2\gamma(\gi-\gamma)}{\gi}\left(\Cov(\EN,\EN_i-B)+\Var[B]\right) \\
\nonumber  &+& \frac{2\gamma(2\gamma-\gi)}{\gi}\Cov(B,\EN-\EN_i).
\end{eqnarray}
Hence, in order to solve problem \eqref{eq:best response problem},
it is enough to find $B\in\ltwo$ that minimizes the quantity
$$\frac{\gi-\gamma}{2\gamma}\Var[\EN_i-B]+\Cov(\EN,\EN_i-B)+\Var[B]+\Cov(B,\EN-\EN_i)\left(1-\frac{\gamma}{\gi-\gamma}\right).$$
Since the above quantity is constant-invariant, we may consider
for now that $\EE[B]=0$. Thus, we have to minimize function
$W_i:\ltwo\rightarrow\R$ defined by
$$W_i(B)=\EE\left[\frac{\gi-\gamma}{2\gamma}(\EN_i-B)^2+\EN(\EN_i-B)+B^2+
\left(1-\frac{\gamma}{\gi-\gamma}\right)B(\EN-\EN_i)\right].$$ The
minimum of $W_i(\cdot)$ can be obtained through its
\textit{Fr\'{e}chet derivative}. For this, we first show that the
Fr\'{e}chet derivative of $W_i(\cdot)$ at $B\in\ltwo$
(denoted by $D_{W_i(B)}$) is given by
$$D_{W_i(B)}[X]=\EE\left[\left(\frac{\gi+\gamma}{\gamma}B
-\frac{\gi-\gamma}{\gamma}\EN_i-\EN+\left(1-\frac{\gamma}{\gi-\gamma}\right)(\EN-\EN_i)\right)X\right],$$
for every $X\in\ltwo$. In order to prove this statement, it
is enough to get the following limit
\begin{equation}\label{eq:best response proof limit}
    \frac{\vert
    W_i(B+X)-W_i(B)-D_{W_i(B)}[X]\vert}{||X||_{\ltwo}}\longrightarrow
    0
\end{equation}
as $||X||_{\ltwo}\rightarrow 0$. Indeed,
$$\frac{\vert W_i(B+X)-W_i(B)-D_{W_i(B)}[X]\vert}{||X||_{\ltwo}}=
\left(\frac{\gi+\gamma}{2\gamma}\right)\frac{\vert\EE\left[X^2\right]\vert}{||X||_{\ltwo}}=
\left(\frac{\gi+\gamma}{2\gamma}\right)||X||_{\ltwo}\longrightarrow
0.$$ Note that function $W_i(\cdot)$ is strictly convex in
$\ltwo$ and hence, if there exists $B^*\in\ltwo$ such
that $D_{W_i(B^*)}[X]=0$ for each $X\in\ltwo$, then $B^*$ is
the unique (in the class of $\ltwo$ with expectation equal to
zero) minimizer of $W_i(\cdot)$. Clearly, for $\Br_i$ given in
\eqref{eq:best response}, $D_{W_i(\Br_i)}[X]=0$ for every
$X\in\ltwo$, which makes $\Br_i$ the best-endowment response of agent
$i$.

\subsection*{Proof of Theorem \ref{thm:nash}}
 From Proposition \ref{pro:best
response} and the FOCs for the Nash equilibrium, we get that Nash
equilibria are the solutions $(\Bn_1,...,\Bn_n)\in(\ltwo)^n$
of the following system of $n$ linear equations
\begin{eqnarray}\label{eq:system}
  \nonumber \Bn_1 &=& \frac{\gamma_1}{\gamma_1+\gamma}\EN_1+\frac{\gamma^2}{\gamma_1^2-\gamma^2}\sum_{j\neq 1}^n \Bn_j \\
  \Bn_2 &=& \frac{\gamma_2}{\gamma_2+\gamma}\EN_2+\frac{\gamma^2}{\gamma_2^2-\gamma^2}\sum_{j\neq 2}^n \Bn_j \\
  \nonumber & \vdots & \\
  \nonumber \Bn_n &=& \frac{\gamma_n}{\gamma_n+\gamma}\EN_n+\frac{\gamma^2}{\gamma_n^2-\gamma^2}\sum_{j=1}^{n-1} \Bn_j.
\end{eqnarray}
We fix $\omega\in\Omega$ and we are looking for
$\mathbf{B}=(B_1,...,B_n)in(\ltwo)^n$ which solves the above linear system.
For each $i\in\{1,...,n\}$, we have that $\Bn_i=
\frac{\gamma_i}{\gamma_i+\gamma}\EN_i+\frac{\gamma^2}{\gamma_i^2-\gamma^2}\sum_{j=1}^n
\Bn_j-\frac{\gamma^2}{\gamma_i^2-\gamma^2}\Bn_i$, which gives
\eqref{eq:B*i} and \eqref{eq:B*}. Since $\Bn_i(\omega)$ is a
linear combination of $\EN_1(\omega),....,\EN_n(\omega)$,
$\Bn_i:\Omega\longrightarrow\R$ is an $\FF$-measurable and, in
particular, it belongs to $\ltwo$.

Finally, it is left to observe that if $\BBn=(\Bn_1,...,\Bn_n)$ is
a Nash equilibrium, every vector of the form $\BBn+\mathbf{c}$,
for $\mathbf{c}\in\R^n$ is also a Nash equilibrium, since it
satisfies \eqref{eq:condition for Nash equilibrium}. The fact that
this form of Nash equilibria is unique follows from the strict
convexity of $G_i(\cdot)$. Equations \eqref{eq:C*} and \eqref{eq:utility gain nash} follow by
straightforward calculations.

\subsection*{Proof of Proposition \ref{pro:n goes to infty}}
We assume without loss of generality that $\EE[\EN_i]=0$ for each
$i\in\N$. The uniform integrability assumption guarantees that there exists a positive constant $M$, such that $||\EN_i||_{\ltwo}\leq M$, for
 all $i\in\N$. It is enough to show that
the first term in \eqref{eq:loss}, i.e., the sum
$\sum_{i=1}^n\gamma_i\Var[\EN_i-\Bn_i]$ vanishes, as $n$ goes to
infinity. We fix $n\in\N$ and for an arbitrarily chosen
$i\in\{1,...,n\}$ we have
\begin{equation}\label{eq:nprof1}
    ||\EN_i-\Bn_i||^2=\left(\frac{\gamma_{-i}}{\gamma_{i}+\gamma_{-i}}\right)^2||\EN_i||^2+
    \left(\frac{\gamma_{-i}}{\gamma_{i}+\gamma_{-i}}\right)^4||\BBn(n)||-
    2\left(\frac{\gamma_{-i}}{\gamma_{i}+\gamma_{-i}}\right)^3\langle\EN_i,\BBn(n)\rangle,
\end{equation}
where $\gamma_{-i}=(\sum_{j\neq i}^n1/\gamma_j)^{-1}$,
$\BBn(n)$ is the aggregated submitted endowment of the $n$ first
agents, all mentioned norms are in $\ltwo$ and
$\langle\cdot,\cdot\rangle$ is the associated inner product.
\begin{eqnarray*}
  \sum_{i=1}^n\gamma_i\left(\frac{\gamma_{-i}}{\gamma_{i}+\gamma_{-i}}\right)^2||\EN_i||^2 &\leq & Mc_u \sum_{i=1}^n\left(\frac{\gamma_{-i}}{\gamma_{i}+\gamma_{-i}}\right)^2\\
   &\leq & \frac{Mc_u^3}{(n-1)^2}\sum_{i=1}^n\frac{1}{\left(\gamma_{i}+\gamma_{-i}\right)^2} \\
   &\leq & \frac{Mc_u^3}{(n-1)^2}\frac{n(n-1)^2}{((n-1)c_l+c_u)^2}
\end{eqnarray*}
which goes to zero as $n\rightarrow \infty$. Also,
\begin{eqnarray*}
  ||\BBn(n)||^2 &=& \frac{1}{1-\sum_{i=1}^n\left(\frac{\gamma(n)}{\gamma_i}\right)^2}\left(||\EN(n)||^2+
    \gamma(n)^2\left\vert\left\vert\sum_{i=1}^n\frac{\EN_i}{\gamma_i}\right\vert\right\vert^2-
    2\gamma(n)\left\langle\EN(n),\sum_{i=1}^n\frac{\EN_i}{\gamma_i}\right\rangle\right) \\
   &\leq & \frac{1}{1-\frac{c_l^2}{nc_u^2}}\left(n^2M+\frac{c_u^2}{n^2}\left\vert\left\vert\sum_{i=1}^n\frac{\EN_i}{\gamma_i}\right\vert\right\vert^2
   +2\frac{c_u}{n}\left\vert\left\langle\EN(n),\sum_{i=1}^n\frac{\EN_i}{\gamma_i}\right\rangle\right\vert\right)
\end{eqnarray*}
where $\gamma(n)$ and $\EN(n)$ stands for the aggregate risk
aversion and endowment of the $n$ first agents. But,
$\left\vert\left\vert\sum_{i=1}^n\frac{\EN_i}{\gamma_i}\right\vert\right\vert^2\leq
\frac{1}{c_l}\left\vert\left\vert\sum_{i=1}^n\EN_i\right\vert\right\vert^2\leq
\frac{n^2M}{c_l}$ and
    $$\left\vert\left\langle\EN(n),\sum_{i=1}^n\frac{\EN_i}{\gamma_i}\right\rangle\right\vert \leq \frac{1}{c_l}\sum_{i=1}^n\left\vert\left\langle\EN(n),\EN_i\right\rangle\right\vert
    \leq
    \frac{1}{c_l}\vert\vert\EN(n)\vert\vert\sum_{i=1}^n\vert\vert\EN_i\vert\vert\leq
    \frac{n^2M}{c_l}.$$
Therefore, we have the following estimation for the aggregated
submitted endowment of the $n$ first agents
\begin{equation}\label{eq:nprof2}
 ||\BBn(n)||^2\leq \frac{M}{c}\left(\frac{cn^2+C^2+2nC}{1-\frac{c^2}{nC^2}}\right).
\end{equation}
Also,
$\sum_{i=1}^n\gamma_i\left(\frac{\gamma_{-i}}{\gamma_{i}+\gamma_{-i}}\right)^4\leq\frac{c_u^5}{c_l^4n^3}$,
which together with \eqref{eq:nprof2} implies that
\begin{equation}\label{eq:nprof3}
\sum_{i=1}^n\gamma_i\left(\frac{\gamma_{-i}}{\gamma_{i}+\gamma_{-i}}\right)^4||\BBn(n)||^2\leq
\frac{Mc_u^5}{c_l^5}\left(\frac{c_ln^2+c_u^2+2nc_u}{n^3-\frac{n^2c_l^2}{c_u^2}}\right)\rightarrow
0
\end{equation} as $n\rightarrow\infty$. Continuing with the terms in
\eqref{eq:nprof1}, we have
\begin{eqnarray}\label{eq:nprof4}
  \sum_{i=1}^n\gamma_i\left(\frac{\gamma_{-i}}{\gamma_{i}+\gamma_{-i}}\right)^3\langle\EN_i,\BBn(n)\rangle &\leq &
  \frac{c_u^4}{c_l^3n^3}\vert\vert\BBn(n)\vert\vert\sum_{i=1}^n\vert\vert\EN_i\vert\vert \nonumber\\
   &\leq & \frac{c_u^4}{c_l^3n^2}\sqrt{M}\vert\vert\BBn(n)\vert\vert \nonumber\\
   &\leq &
   \frac{c_u^4}{c_l^{7/2}n^2}M\sqrt{\frac{c_ln^2+c_u^2+2nc_u}{1-\frac{c_l^2}{nc_u^2}}}\rightarrow 0
\end{eqnarray}
as $n\rightarrow\infty$. From \eqref{eq:nprof1}, \eqref{eq:nprof3}
and \eqref{eq:nprof4} we get that the first term of
\eqref{eq:loss} goes to zero as the number of agents approaches
infinity.

\subsection*{Proof of Theorem \ref{thm:Nash equilibrium price}}
Taking Proposition \ref{pro:best price response} into account, we
conclude that the market equilibrates when the covariances
$\Cov(\CC,\Bn_i)$ equilibrate. As we have seen in Theorem
\ref{thm:nash}, this may happen for endowments $\Bn_i$ given in
\eqref{eq:B*i}, which then yields price \eqref{eq:Nash equilibrium price of C}. The uniqueness of the Nash equilibrium price follows by the
Standing Assumption. 

The equivalence of Nash and the competitive 
equilibrium prices when agents have common risk aversion is then
induced by Corollary \ref{cor:sharing game} (see also equation
\eqref{eq:PEquadr}). Finally, \eqref{eq: Nash equilbirium allocation} follows by Proposition \ref{pro:equilibrium with
quadratic}, and simple calculations give \eqref{eq: Nash level C}.

\subsection*{Proof of Proposition \ref{pro:nC goes to infty}}
We assume as usual that expectations are equal to zero and from
\eqref{eq:PEquadr} and \eqref{eq: Nash equilbirium allocation} we
have that
$$||\pp^*(n)-\ppn(n)||=2\gamma(n)\vert\vert\Cov(\CC,\BBn(n)-\EN(n))\vert\vert
\leq
2\gamma(n)\vert\vert\BBn(n)-\EN(n)\vert\vert\sqrt{\sum_{j=1}^k\vert\vert
C_j\vert\vert^2},$$ where the notation is the one introduced in
the proof of Proposition \ref{pro:n goes to infty}. Therefore, it
is enough to show that
$\gamma(n)\vert\vert\BBn(n)-\EN(n)\vert\vert$ vanishes as
$n\rightarrow\infty$. For sufficiently large $n$, we have
\begin{eqnarray*}
  \gamma(n)\vert\vert\BBn(n)-\EN(n)\vert\vert &=& \frac{\gamma^2(n)}{1-\sum_{i=1}^n\left(\frac{\gamma(n)}{\gamma_i}\right)^2}
  \left\vert\left\vert\sum_{i=1}^n\frac{\gamma_i\EN_i-\gamma(n)\EN(n)}{\gamma_i^2}\right\vert\right\vert \\
   &\leq & \frac{\gamma^2(n)}{c_l^2-\frac{c_u^2}{n}}\left\vert\left\vert n\gamma(n)\EN(n)-\sum_{i=1}^n\gamma_i\EN_i\right\vert\right\vert
   \leq \frac{c_u^2}{n^2c_l^2-nc_u^2}\left(n\gamma(n)\left\vert\left\vert\EN(n)\right\vert\right\vert+\left\vert\left\vert\sum_{i=1}^n\gamma_i\EN_i\right\vert\right\vert\right)  \\
   &\leq &
   \frac{\sqrt{M}c_u^2}{n^2c_l^2-nc_u^2}\left(n^2\gamma(n)+nc_u\right)\leq
   2\frac{n\sqrt{M}c_u^3}{n^2c_l^2-nc_u^2}\rightarrow 0
\end{eqnarray*}
as $n\rightarrow\infty$. For item (ii), we fix an agent $i$ and
from \eqref{eq: optimal contracts} and \eqref{eq: Nash equilbirium
allocation}, it is enough to show that
$\vert\vert\Cov(\CC,\Cn_i-C_i^o)\vert\vert$ goes to zero as
$n\rightarrow\infty$. Hence, it suffices to
show that $\vert\vert \Cn_i-C_i^o\vert\vert\rightarrow 0$.
\begin{eqnarray*}
  \vert\vert \Cn_i-C_i^o\vert\vert &=& \frac{\gamma(n)}{\gamma_i}\left\vert\left\vert\EN_i+\frac{\gamma_i-\gamma(n)}{\gamma_i}\BBn(n)-\EN(n)\right\vert\right\vert \\
   &\leq & \frac{\gamma(n)}{\gamma_i}\vert\vert \EN_i\vert\vert+
   \frac{\gamma(n)}{\gamma_i}\vert\vert\BBn(n)-\EN(n)\vert\vert+   \frac{\gamma^2(n)}{\gamma^2_i}\vert\vert\BBn(n)\vert\vert.
\end{eqnarray*}
We have seen in the proof of item (i) that
$\gamma(n)\vert\vert\BBn(n)-\EN(n)\vert\vert\rightarrow 0$. Also,
since $\gamma(n)\leq\frac{c_u}{n}$,
$\frac{\gamma(n)}{\gamma_i}\vert\vert \EN_i\vert\vert\rightarrow
0$. Finally, taking into account \eqref{eq:nprof2}, we get that
$\gamma^2(n)\vert\vert\BBn(n)\vert\vert\rightarrow 0$, which
completes the proof.

\subsection*{On proofs of Sections \ref{sec: noise trader} and \ref{sec:difference}}\

The proofs of results stated in these sections follow the arguments developed in the previous ones. For that reason, we skip the repetition of details
and mention only the main differences in the proofs. 

For the proof of Proposition \ref{pro:n Pareto}, we follow the arguments of the proof of Proposition \ref{pro:optimal sharing}; the Pareto-optimal securities are the one¦Ò that maximize the sum of agents' utility functionals under the set $\tilde{\A}$. For this we similarly show that payoffs $(\nC_i^o)_{i=1}^n$ given in \eqref{eq:n optimal securities} maximize the sum $\sum_{i=1}^n\UU_i(\EN_i+C_i)$, while the equilibrium prices for each security $\nC_i^o$ is given as a special case of Proposition \ref{pro:optimal sharing}.
\smallskip

For proof of Theorem \ref{thm:n nash}, we first get that the FOCs for the Nash equilibrium endowments $(\nBn_1,...,\nBn_n)\in\ltwo$ are the solution of the system of equations $\nBn_i = \gamma_i\EN_i/(\gamma_i+\gamma)+\gamma^2(\sum_{j\neq i}^n \Bn_j+\EN_N)/(\gamma_i^2-\gamma^2)$, for each $i\in\{1,...,n\}$ (see also \eqref{eq:system}). This system is solved omega-wise, and gives the solution stated in \eqref{eq:n B*i} and \eqref{eq:nB*}. Uniqueness (modulo constants) follows as in Theorem \ref{thm:nash}.
\smallskip
Regarding the Proposition \ref{pro:n best response ini}, the main difference with Propositions \ref{pro:best price response} and \ref{pro: best demand and endowment} is that agent $i$ responds to aggregate demand $\left[(\EE[\CC]-\pp)/2\gamma_{-i}-\Cov(\CC,\EN_{-i}-\zeta\cdot\CC)\right]\cdot\Var^{-1}[\CC]$, which is in fact the demand function induced by the endowment $\EN_{-i}-\zeta\cdot\CC$. In the same spirit, in order to prove Theorem \ref{thm:n Nash equilibrium price}, we recall the proof of Theorems \ref{thm:Nash equilibrium price} and \ref{thm:n nash}, where the noise traders' endowment is equal to $-\zeta\cdot\CC$. 
\smallskip
The limiting argument in Corollary \ref{cor:n gains incomplete} follows by straightforward calculations thanks to the fact that $\lzer$-$\lim_{\gamma_1\rightarrow 0}C_1^o=\EN_2$.  
\bigskip

For the proof of Proposition \ref{pro:df best-response price and demand}, we need to verify that price vector $\dppr_i$ given in \eqref{eq:d best price}
is the unique maximizer of the function $\R^k\ni\pp\mapsto\UU_i(\EN_i-Z_{-i}(\pp)\cdot\CC)+Z_{-i}(\pp)\cdot\pp$. Indeed, this is a strictly convex function with derivative equal to zero at price vector $\dppr_i$, and the representation of the best-demand response $\dZr_i$ readily follows.
\smallskip 
Based on results of Proposition \ref{pro:df best-response price and demand}, we get that the conditions for Nash equilibrium are equivalent with a simple linear system of equations for the intercept points $(\inin_i)_{i=1}^n$, given by $\inin_i=\inin_i\lambda^2_i/(1-\lambda^2_i)+\ini_i/(1+\lambda_i)$, for each $i$. Its solution is given in \eqref{eq:db Nash demands ini}, a fact that finishes the proof of Theorem \ref{thm:db Nash equilibrium price}.
\smallskip
Finally, Corollary \ref{cor:db homo} readily follows since $\alpha=(n-1)/n$ when agents are homogeneous. Note also that, as pointed out in Remark \ref{rem:db homo}, equality $\dpp=\dppn$ does not necessarily imply that $\gamma_i$'s are the same, since there exists a combination of different beliefs and different risk aversions that gives equality $\ini=\inin$.

\section{An Example of Restricted Strategic Sets}\label{sec:best response with percentage}

The best-endowment response \eqref{eq:best response} and the corresponding Nash equilibrium assume that the agents' set of strategic choices is whole $\ltwo$. This means that an agent can report (almost) any random variable as her endowment. In fact, it is shown that she has motive to report exposure to the other agents' endowment. An indirect outcome of this possibility is that agents with sufficiently lower risk aversion benefit from Nash equilibrium transaction. 

Under a more practical perspective however, an agent may be concerned
that revealing exposure to the risks that only the other agents
have, will deteriorate the conditions of future businesses with
them. This essentially means that the set of strategic choice may be exogenously restricted$^{38}$\footnote{$^{38}$In some of the risk-sharing transactions, the participating agents have endowments of special type. For example, insurance companies may want to share insurance contracts on specific catastrophic events, or large short positions on CDS written on specific reference entities.}. It is worth examining how the Nash equilibrium is affected when restrictions on the strategic set are imposed and whether low risk averse agents still get more utility at Nash equilibrium. Although the detailed analysis of this problem could be a subject of a separate study, we present here a simple but indicative example, where agents exploit their market power by choosing (not the endowments they submit but rather)
the size of their true risk exposures that are going to submit for risk sharing.$^{39}$\footnote{$^{39}$In this restricted problem, it is assumed that each agent knows the direction of the other agents' risk exposure, but not the exact sizes. For instance, in the reinsurance market, it is publicly known that some insurance companies have taken several positions on specific contracts, but none knows their exact net sizes.} 
In other words, the agents' set of strategic choices is
\textit{restricted} to positive proportions (possibly higher than one) of the true
endowment. To emphasize the difference with problem \eqref{eq:best response
problem}, we use the notation
$g_i(b;\EN_{-i}):=G_i(b\EN_i;\EN_{-i})$, and hence problem
 \eqref{eq:best response problem} becomes
\begin{equation}\label{eq:best response percentage problem}
    \br_i:=\underset{b\in [0,\infty)}{\argmax}g_i(b;\EN_{-i})
\end{equation}
where $\br_i$ is called \textit{best-percentage
response}. We deal with this restriction in the proposition below,
the proof of which follows the lines of Proposition
\ref{pro:best response}.

\begin{proposition}\label{pro:best response percentage}
For each $i\in\{1,...,n\}$, the unique best-percentage response of agent $i$, when the rest of the agents have
reported aggregate endowment $\EN_{-i}$, is given by
\begin{equation}\label{eq:best response percentage of agent-i}
    \br_i=0\vee\left(\frac{1}{1+\lambda_i}+\frac{\lambda_i^2}{1-\lambda_i^2}\rho(\EN_i,\EN_{-i})\sqrt{\frac{\Var[\EN_{-i}]}{\Var[\EN_i]}}\right),
\end{equation}
where $\rho(.,.)$ is the correlation coefficient map.
\end{proposition}


As in the best-endowment response problem, it is (almost) never optimal for an agent to report the true size of her risk exposure. In fact, the best-percentage response is an increasing function of $\rho\left(\EN_{i},\EN_{-i}\right)$, and agent $i$ takes a speculation-only position (i.e., she buys a part of other agents' endowment and short none of her true endowment), if the correlation is sufficiently below zero. This behavior is at first glance
    surprising, since negative endowments' correlation implies good
    hedging. However, one of the goals of this predatory trading is to exploit the knowledge on the other agents' endowment. If
    this endowment is negatively correlated with $\EN_i$, it is
    preferable for agent $i$ to share less of her risk exposure in order to achieve a
    better cash compensation for buying some of the other agents'
    risk and at the same time exploit the negative
    correlation for her own true hedging needs. The
    situation differs when $\rho\left(\EN_i,\EN_{-i}\right)$ is positive. When
    the correlation is close to one, agent $i$ may report
    overexposure on $\EN_i$. This implies that after the transaction she will be left
    with less exposure in $\EN_i$, which together with the simultaneous long position on $\lambda_i\EN_{-i}$,
    results in a total position of less variance.

Under this restricted strategic set, the Nash risk-sharing equilibrium is defined below.
\begin{definition}\label{def: Nash equilibrium percentage}
We call a vector $(\bn_1,...,\bn_n)\in[0,\infty)^n$ restricted
Nash risk-sharing equilibrium if for each $i\in\{1,...,n\}$
\begin{equation}\label{eq:condition for percentage Nash equilibrium}
    g_i(\bn_i;\sum_{j\neq i}\bn_j\EN_{j})\geq g_i(b;\sum_{j\neq i}\bn_j\EN_{j}),\quad\quad\text{for all } b\in[0,\infty).
\end{equation}
\end{definition}
Note that the induced Nash risk-sharing securities are given by $\Cn_i(\bn_i)=\lambda_i\sum_{j\neq i}\bn_j\EN_{j}-\lambda_{-i}\bn_i\EN_i$, for each $i\in\{1,...,n\}$. Thanks to Glicksburg-Fan-Debreu Theorem (see among others Chapter
1 of Fudenberg and Tilore \cite{FunTir91}) and the linearity of the best-response
function for positive values (see \eqref{eq:best response
percentage of agent-i}), we are able to establish the existence
and the uniqueness of this equilibrium, provided there is an
exogenously given bound on agents' choices.$^{40}$\footnote{$^{40}$This bound in the set of choices
mitigates the magnitude of the overexposure of the reported endowments, however it is
by no means restrictive when we consider real-world situations.}
\begin{proposition}\label{pro:Nash equilibrium percentage}
Suppose that the set of choices for each agent is bounded from
above by an upper bound $\kappa>0$. Then, there exists a unique
restricted Nash risk-sharing equilibrium.
\end{proposition}

In the simplified case of $n=2$, Nash risk-sharing
equilibrium, $(\bn_1,\bn_2)$ solves the equations
\begin{equation}\label{eq:equilibrium percentage}
\bn_i=\left(0\vee\left(\frac{1}{1+\lambda_i}+\frac{\lambda_i^2\bn_{-i}}{\lambda_{-i}(1+\lambda_i)}\rho(\EN_{i},\EN_{-i})\sqrt{\frac{\Var[\EN_{-i}]}{\Var[\EN_i]}}\right)\right)\wedge\kappa,
\,\,\,\,\,\text{ for }i=1,2.^{41}\footnote{$^{41}$In the case
where $\EN_i$ is constant,
$\bn_i=\frac{1}{1+\lambda_i}\wedge\kappa$.}
\end{equation}

\begin{figure}[!ht]
\includegraphics[trim = 35mm 0mm 0mm 0mm, clip, scale=0.55]{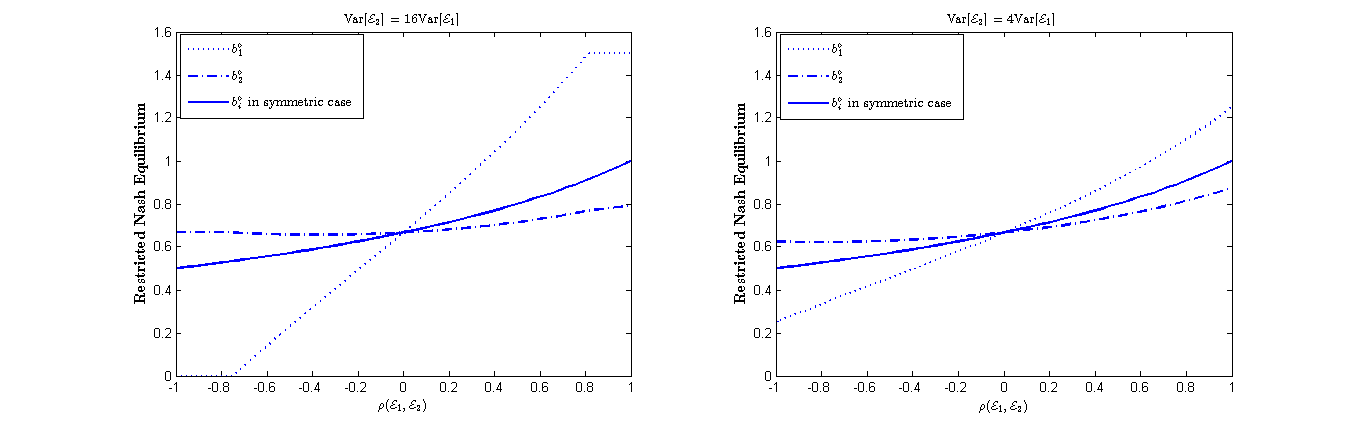}
\caption{{\footnotesize Restricted Nash Equilibria with $\gamma_1=\gamma_2$ and $k=1.5$. The symmetric case refers to the equality $\Var[\EN_1]=\Var[\EN_2]$}} \label{fig:nash}
\end{figure}

The equilibrium's dependence on the endowments' correlation can be
isolated if agents are homogeneous with respect to their risk aversion. Three different choices of the fraction
$\Var[\EN_2]/\Var[\EN_1]$ are illustrated in Figure \ref{fig:nash}. As expected, the agent with riskier endowment (agent
2 in this particular example) implements the strategic behavior less
intensely, since her increased hedging needs count more than the
cash transfer (all else equal). 

Regarding the risk-aversion coefficients, $\bn_i$ is a decreasing
(resp.~increasing) function of $\gamma_i$ when $\rho(\EN_1,\EN_2)$ is
positive (resp.~negative), implying that high risk averse agents are
more reluctant to apply this strategic behavior; in the extreme
case it holds that $\underset{\gamma_i\rightarrow
\infty}{\lim}\bn_{i}=1$. Furthermore, as in the unrestricted case,
there is a similar asymptotic interaction between agents' behavior
and their risk aversions. Namely,
\begin{center}
$\underset{\gamma_i\rightarrow \infty}{\lim}\bn_{-i}=\underset{\gamma_{-i}\rightarrow 0}{\lim}\bn_{-i}=\left\{%
\begin{array}{ll}
    0, & \hbox{$\rho\left(\EN_i,\EN_{-i}\right)<0$;} \\
    \frac{1}{2}, & \hbox{$\rho\left(\EN_i,\EN_{-i}\right)=0$;} \\
     \kappa, & \hbox{$\rho\left(\EN_i,\EN_{-i}\right)>0$,} \\
\end{array}%
\right.\,\,\,\,\,\,$ and
$\,\,\,\,\,\,\underset{\gamma_i\rightarrow
\infty}{\lim}\bn_{i}=\underset{\gamma_{i}\rightarrow
0}{\lim}\bn_{-i}=1$.
\end{center}
The above limits verify that even in this restricted game, an
agent with risk preferences close to risk neutrality dominates the
risk-sharing transaction. However, in contrast to the unrestricted game, it is not necessarily true that low risk averse agents get more utility at the Nash game than at the optimal risk-sharing transaction. One simple counterexample is the case where $n=2$, $\kappa\geq 1$ and $\rho(\EN_1,\EN_2)=0$, which together with \eqref{eq:equilibrium percentage} imply that $\bn_i=1/(1+\lambda_i)$, for both $i=1,2$. Then, when $\Var[\EN_1]/\Var[\EN_2]$ is sufficiently small, Nash equilibrium is always worse than Pareto for agent 1, for all $\gamma_1>0$. In fact, we readily observe that the difference of agent's utility at the Pareto-optimal risk-sharing and at the restricted Nash risk-sharing equilibrium converges to zero as $\gamma_1\downarrow 0$. This counterexample highlights the importance of agents' ability to submit exposure to the other agents' endowment regarding the level of agents' utility surpluses.


\bigskip

\begin{thebibliography}{53}
\bibitem{Acc07}
Acciaio B.: Optimal risk sharing with non-monotone monetary
functions. Finance and Stochastics 11 (2) 267--289 (2007)

\bibitem{AchaBin05}
Acharya V.V., Bisin, A.: Optimal financial-market integration and
security design. Journal of Business, vol. 78, no. 6, 2397--2433
(2005)
\bibitem{AllGal94}
Allen, F., Gale, D.: \textit{Financial Innovation and Risk
Sharing}. Cambridge, MA: MIT press  (1994)

\bibitem{AnthKar15}
Anthropelos, M., Kardaras, C.: Equilibrium in risk-sharing games. Preprint available at \text{//arxiv.org/abs/1412.4208} (2015)

\bibitem{AnthZit10a}
Anthropelos, M., \v{Z}itkovi\'{c}, G.: On agents' agreement and
partial equilibrium pricing in incomplete markets. Mathematical
Finance, 20, no. 3, 411--446 (2010)


\bibitem{BacCaoWil00}
Back, K., Cao, C.H., Willard, G.A.: Imperfect competition among informed traders. Journal of Finance, 55, 2117--2155 (2000)

\bibitem{ElKBarr05}
Barrieu, P., EL Karoui, N.: Inf-convolution of risk measures and
optimal risk transfer. Finance and Stochastics, 9, no. 2, 269--298
(2005)

\bibitem{Bor62}
Borch, K.: Equilibrium in reinsurance market. Econometrica, 30,
424--444 (1962)

\bibitem{Bor68}
Borch, K.: General equilibrium in the economics of uncertainty. In
\textit{Risk and Uncertainy}, K. Borch and J. Mossin eds.,
Macmillan, London (1968)

\bibitem{Bru01}
Brunnermeier, M.: \textit{Asset Pricing under Asymmetric Information - Bubbles, Crashes,
Technical Analysis and Herding}. Oxford University Press (2003)

\bibitem{BruPed05}
Brunnermeier, M, Pedersen, L.H.: Predatory trading. Journal of
Finance, 60, 1825--1863 (2005)

\bibitem{CarEkTou07}
Carlier, G., Ekeland, I., Touzi, N.: Optimal derivatives design
for mean-variance agents under adverse selection. Mathematics and
Financial Economics, 1, 57--80  (2007)

\bibitem{Car14}
Carvajal, A.: Arbitrage pricing in non-competitive financial
markets. Working paper (2014)

\bibitem{CarRosWer12}
Carvajal, A., Rostek, M., Weretka, M.: Competition in financial innovation. Econometrica, 80, 1895--1936 (2012)

\bibitem{CarWer12}
Carvajal, A., Weretka, M.: No-arbitrage, state prices and trade in
thin financial markets. Economic Theory, 50, 223--268 (2012)


\bibitem{ChaLak95}
Chan, L., Lakonishok, J.: The behavior of stock price around
institutional trades. Journal of Finance, 50, 1147--1174  (1995)

\bibitem{ChrHarSch94}
Christie, W., Harris, J., Schultz, P.: Why did NASDAQ market
makers stop avoiding odd-eight quotes? Journal of Finance, 49,
1841--1860 (1994)

\bibitem{ChrSch94}
Christie, W., Schultz, P.: Why did NASDAQ market makers avoid
odd-eight quotes? Journal of Finance, 49, 1841--1860 (1994)


\bibitem{Dan11}
Dana, R.-A.: Comonotonicity, efficient risk-sharing and equilibria
in markets with short selling for concave law-invariant utilities.
Journal of Mathematical Economics 47, 328--335 (2011)

\bibitem{DemLar95}
Demange, G., Laroque, G.: Optimality of incomplete markets.
Journal of Economic Theory, 65, 218--232 (1995)

\bibitem{DufGArPed05}
Duffie, D., Garleanu, N., Pedersen, L.H.: Over-the-counter markets. Econometrica 73, 1815¨C-1847 (2005)

\bibitem{DufGArPed07}
Duffie, D., Garleanu, N., Pedersen, L.H.: Valuation in Over-the-counter markets. Review of Financial Studies 20 1865¨C-1900 (2007)

\bibitem{DuffRah95}
Duffie, D., Rahi, R.: Financial market innovation and security
design: An introduction. Journal of Economic Theory, 65 (1), 1--42
(1995)

\bibitem{FunTir91}
Fudenberg, D., Tirole, J.: \textit{Game Theory}, The MIT Press
(1991)

\bibitem{GibSinYer03}
Gibson, S. Singh, R. Yerramilli, V.: The effect of decimalization
on the components of the bid-ask spread. Journal of Financial
Intermediation, 12, 121--148 (2003)

\bibitem{Gro81}
Grossman, S.: Nash equilibrium and the industrial organization of
markets with large fixed costs. Econometrica, 49, 1149--1172
(1981)

\bibitem{Hel80}
Hellwig, M.R.: On the aggregation of information in competitive
markets. Journal of Economic Theory, 22, 477--498 (1980)

\bibitem{HorMorBro08}
Horst, U., Moreno-Bromberg, S.: Risk minimization and optimal
derivative design in a principal agent game. Mathematics and
Financial Economics 2, (1), 1--27 (2008)

\bibitem{HorMorBro11}
Horst, U., Moreno-Bromberg, S.: Efficiency and equilibria in games
of optimal derivative design. Mathematics and Financial Economics,
5, 269--297 (2011)

\bibitem{JouSchTou08}
Jouini, E., Schachermayer, W., Touzi N.: Optimal risk sharing for
law invariant monetary utility functions. Mathematical Finance, 2
vol. 18, 269--292 (2008)

\bibitem{KeiMad95}
Keim, D., Madhavan, A.: Anatomy of the trading process: Empirical
evidence on the motivation for and execution of institutional
equity trades. Journal of Financial Economics, 37, 371--398 (1995)

\bibitem{KeiMad97}
Keim, D., Madhavan, A.:Transactions costs and investment styles:
An inter-exchange analysis of institutional equity trades. Journal
of Financial Economics, 46, 265--292 (1997)

\bibitem{KleMey89}
Klemperer, P., Meyer, M.A.: Supply function equilibria in
oligopoly under uncertainty. Econometrica, 57, 1243--1277 (1989)

\bibitem{Kou03}
Koutsougeras, L.: Non-Walrasian equilibria and the law of one
price. Journal of Economic Theory, 108, 169-"1¤7175 (2003)

\bibitem{KouPap04}
Koutsougeras, L., Papadopoulos, K.: Arbitrage and equilibrium in
strategic security markets. Economic Theory 23, 553-"1¤7568 (2004)

\bibitem{KraSto72}
Kraus, A., Stoll, H.: Price impact on block trading on the New
York stock exchange. Journal of Finance, 27, 569--588 (1972)

\bibitem{Kyle89}
Kyle, A.S.: Informed speculation with imperfect competition. The
Review of Economic Studies, 56, 317--356 (1989)

\bibitem{KylObiWan14}
Kyle, A.S., Obizhaeva, A.A., Wang, Y.: Smooth trading with
overconfidence and market power. Preprint available at
\textit{http://papers.ssrn.com/sol3papers.cfm/abstract id=2423207}
(2014)

\bibitem{LiWan16}
Liu, H, Wang, Y.: Market making with asymmetric information and inventory risk. Journal of Economic Theory, 163, 73--109 (2016)

\bibitem{MagQui96}
Magill, M., Quinzii, M.: \textit{Theory of Incomplete Markets},
Vol.1, MIT Press (1996)


\bibitem{MalRos15}
Malamud, S., Rostek, M.: Decentralized exchange. Working Paper (2015)

\bibitem{PagMont03}
Page, F.Jr., Monteiro, P.: Three principles of competitive
nonlinear pricing, Journal of Mathematical Economics, 39, Issues
1-2, 63--109 (2003)

\bibitem{PagMont07}
Page, F.Jr., Monteiro, P.: Uniform payoff security and Nash
equilibrium in compact games, Journal of Economic Theory, 134,
566--575 (2007)

\bibitem{Pal96}
Palomino, F.: Noise trading in small markets. Journal of Finance,
vol. 51, no. 4, 1537--1550 (1996)

\bibitem{PecShel96}
Peck, J., Shell, K.: On the non-equivalence of the
Arrow-securities game and the contingent commodities game. In
Economic complexity: chaos, sunspots, bubbles, and non-linearity,
J. G. William Barnett and K. Shell, Eds. Cambridge University
Press, 61--85 (1996)

\bibitem{RahZig09}
Rahi, R., Zigrand, J.-P.: Strategic financial innovation in
segmented markets. The Review of Financial Studies, vol. 22, no. 8
2941--2971 (2009)

\bibitem{RosWer12}
Rostek, M., Weretka, M.: Price inference in small markets. Econometrica, vol. 80, No. 2, 687--711 (2012)

\bibitem{RosWer15}
Rostek, M., Weretka, M.: Dynamic thin markets. Forthcoming in The Review of Financial Studies.
(2015)

\bibitem{ShaShu77}
Shapley, L., and Shubik, M.: Trade using one commodity as a means
of payment. Journal of Political Economy 85, 937"1¤7-968 (1977)

\bibitem{ShlVis97}
Shleifer, A., Vishny, R.W.: The limits of arbitrage The Journal of
Finance, 52, No. 1, 35--55 (1997)

\bibitem{Vay99}
Vayanos, D.: Strategic trading and welfare in dynamic way. The
Review of Economic Studies, 66, 219--254 (1999)

\bibitem{Vay01}
Vayanos, D.: Strategic trading in a dynamic noisy market, Journal
of Finance, 2001, 56, 131-171 (2001)

\bibitem{Wer11}
Weretka, M.: Endogenous market power. Journal of Economic Theory,
146, 2281--2306 (2011)

\bibitem{Wil68}
Wilson, R.: The theory of syndicates. Econometrica, 36,
119--132 (1968)



\end{thebibliography}
\end{document}